\documentclass[AMA,STIX1COL]{WileyNJD-v2}

\articletype{Research Article}%

\received{26 April 2022}
\revised{6 June 2022}
\accepted{6 June 2022}

\newcommand\independent{\protect\mathpalette{\protect\independenT}{\perp}}
\def\independenT#1#2{\mathrel{\rlap{$#1#2$}\mkern2mu{#1#2}}} 

\newcommand{\var}{\mbox{var}}

\newcommand{\Tf}{\mathcal{T}_F}
\newcommand{\pr}{\mbox{pr}}
\newcommand{\logit}{\mbox{logit}}
\newcommand{\expit}{\mbox{expit}}
\newcommand{\calM}{\mathcal{M}}
\newcommand{\calD}{\mathcal{D}}
\newcommand{\Lbar}{\bar{L}}
\newcommand{\hatalpha}{\widehat{\alpha}}
\newcommand{\hatbeta}{\widehat{\beta}}

\newcommand{\hatmu}{\widehat{\mu}}

\newcommand{\tilM}{\widetilde{M}}

\newcommand{\calK}{\mathcal{K}}
\newcommand{\calG}{\mathcal{G}}
\newcommand{\hatK}{\widehat{\calK}}

\newcommand{\sumin}{\sum^n_{i=1}}

\newcommand{\hatpsi}{\widehat{\psi}}
\newcommand{\hatpi}{\widehat{\pi}}
\newcommand{\hatphi}{\widehat{\phi}}

\newcommand{\calO}[1]{\mathcal{O}^{(#1)}}
\newcommand{\calF}{\mathcal{F}}
\newcommand{\hatG}{\widehat{\calG}}
\newcommand{\hatY}{\widehat{\mathfrak{Y}}}

\newcommand{\Mc}[1]{M^{(#1)}_c}
\newcommand{\hatMci}[1]{\widehat{M}^{(#1)}_{ci}}
\newcommand{\Lamc}[1]{\Lambda^{(#1)}_c}
\newcommand{\hatLamc}[1]{\widehat{\Lambda}^{(#1)}_c}
\newcommand{\Nc}[1]{N^{(#1)}_c}
\newcommand{\Nci}[1]{N^{(#1)}_{ci}}
\newcommand{\Yt}[1]{\mathcal{Y}^{(#1)}}

\newcommand{\tilNc}[1]{\widetilde{N}^{(#1)}_c}
\newcommand{\tilYt}[1]{\widetilde{\mathcal{Y}}^{(#1)}}

\begin{document}

\title{Group sequential methods for interim monitoring of randomized clinical trials with
  time-lagged outcome}

\author[1]{Anastasios A. Tsiatis}

\author[1]{Marie Davidian*}

\authormark{TSIATIS and DAVIDIAN}

\address[1]{\orgdiv{Department of Statistics}, \orgname{North Carolina State University}, \orgaddress{\state{North Carolina}, \country{USA}}}


\corres{*Marie Davidian, Department of Statistics, North Carolina State University, Raleigh, NC 27695-8203, USA. \email{davidian@ncsu.edu}}


\abstract[Summary]{The primary analysis in two-arm clinical trials
  usually involves inference on a scalar treatment effect parameter;
  e.g., depending on the outcome, the difference of treatment-specific
  means, risk difference, risk ratio, or odds ratio.  Most clinical
  trials are monitored for the possibility of early stopping.  Because
  ordinarily the outcome on any given subject can be ascertained only
  after some time lag, at the time of an interim analysis, among the
  subjects already enrolled, the outcome is known for only a subset
  and is effectively censored for those who have not been enrolled
  sufficiently long for it to be observed.  Typically, the interim
  analysis is based only on the data from subjects for whom the
  outcome has been ascertained.  A goal of an interim analysis is to
  stop the trial as soon as the evidence is strong enough to do so,
  suggesting that the analysis ideally should make the most efficient
  use of all available data, thus including information on censoring
  as well as other baseline and time-dependent covariates in a
  principled way.  A general group sequential framework is proposed
  for clinical trials with a time-lagged outcome.  Treatment effect
  estimators that take account of censoring and incorporate covariate
  information at an interim analysis are derived using semiparametric
  theory and are demonstrated to lead to stronger evidence for early
  stopping than standard approaches.  The associated test statistics
  are shown to have the independent increments structure, so that
  standard software can be used to obtain stopping boundaries.}

\keywords{augmented inverse probability weighting, early stopping, influence
  function, proportion of information}


\maketitle

\section{Introduction}\label{s:intro}

In many randomized clinical trials, the primary analysis involves a
comparison of two treatments, 
typically an active or experimental agent versus a control,
which is formalized as inference on a scalar treatment effect parameter.
When the primary outcome is a continuous measure, this parameter is
usually the difference of treatment-specific means.  For a binary
outcome, the treatment effect parameter may be the risk difference,
risk ratio, or odds ratio; and the odds ratio under the assumption of
a proportional odds model is often the treatment effect parameter of
interest in trials involving an ordinal categorical outcome.  The
primary analysis is ordinarily based on a test statistic constructed using
an estimator for the parameter of interest, e.g., the difference of
sample means or maximum likelihood (ML) estimator for the odds ratio in a
proportional odds model.  The overall sample size is established so
that the power to detect a clinically meaningful departure from the
hypothesis of no treatment effect at a given level of significance
using the test statistic at the final analysis achieves some desired
value, e.g., 90\%.

Most later-stage clinical trials are monitored for the possibility of
early stopping for efficacy or futility by a data and safety
monitoring board (DSMB), with interim analyses planned at either
fixed, predetermined analysis times or when specified proportions of
the total ``statistical information'' to be gained from the completed
trial have accrued\cite{TsiatisIBM}.  Ordinarily, at the time of an
interim analysis, the test statistic to be used for the final analysis
is computed based on the available data and compared to a suitable
stopping boundary constructed to preserve the overall operating
characteristics of the trial\cite{Pocock,OBF}.

Because of staggered entry into the trial, the data available at the
time of an interim analysis are from subjects who have already
enrolled.  Moreover, the primary outcome $Y$ is ordinarily not known
immediately but is ascertained after some lag time $T$, say.  In some
trials, the lag time $T$ is the same for all participants, as in the
case where $Y$ is a continuous outcome that will be measured at a
prespecified follow-up time $\Tf$, e.g., $\Tf=$ one year, so that
$T=\Tf$ for all subjects, and the treatment parameter is the
difference in treatment means of $Y$ at one year.  Here, at the time
of an interim analysis, $Y$ will be available only for subjects
enrolled for at least one year, so that the analysis can be based only
on the data for these subjects.

In other settings, the time lag $T$ may be different for different
participants, as is the case in many clinical trials of COVID-19
therapeutics conducted by the Accelerating COVID-19 Therapeutic
Interventions and Vaccines (ACTIV) public-private partnership.  In an
ongoing clinical trial coordinated through the ACTIV-3b: Therapeutics
for Severely Ill Inpatients with COVID-19 (TESICO) master
protocol\cite{TESICO}, patients hospitalized with acute respiratory
distress syndrome are randomized to receive an active agent or placebo
and followed for up to $\Tf=90$ days.  The primary outcome $Y$ is an
ordinal categorical variable with six levels.  The first five
categories reflect a subject's status at 90 days following enrollment:
1, at home and off oxygen for at least 77 days (the most favorable
category); 2, at home and off oxygen for at least 49 but no more than
76 days; 3, at home and off oxygen for at least 1 but no more than 48
days; 4, not hospitalized and either at home on oxygen or receiving
care elsewhere; and 5, still hospitalized or in hospice care.
Category 6 (the worst) corresponds to death within the 90 day
follow-up period.  While Categories 1-5 cannot be ascertained until a
subject has been followed for the full 90 days, that a subject's
outcome is Category 6 is known at the time of death.  Thus, the time
lag before ascertainment is $T=\Tf = 90$ days for subjects with
$Y = 1,\ldots,5$ and is equal to the (random) time of death
$T\leq \Tf = 90$ for those with $Y=6$.  In TESICO, the treatment
effect parameter is the odds ratio for active agent relative to
placebo under an assumed proportional odds model.  Similarly, in a
clinical trial coordinated through the ACTIV-2: A Study for
Outpatients With COVID-19 master protocol (Study A5401)\cite{ACTIV2},
subjects within seven days of self-reported COVID-19 onset are
randomized to receive an active agent or placebo and followed for up
to $\Tf=28$ days for the binary outcome $Y$, where $Y=1$ if the
subject dies or is or hospitalized within 28 days and $Y=0$ otherwise.
For subjects who die or are hospitalized at time $T$ prior to 28 days,
$Y=1$ is ascertained after a time lag $T \leq \Tf= 28$, whereas that
$Y=0$ can be ascertained only after the full 28 days, and $T=\Tf=28$.
Here, the treatment parameter is the relative risk (risk ratio) of
hospitalization/death for active agent versus placebo.

At the time of an interim analysis in TESICO and A5401, the available
data include the outcomes for all enrolled subjects who have been
followed for at least 90 and 28 days, so for whom $T=\Tf = 90$ or 28,
respectively, along with the outcomes for enrolled subjects who do not
have 90 or 28 days of follow up but have already been observed to die
($Y=6$) in TESICO or to be hospitalized or die ($Y=1$) in A5401
($T \leq \Tf = 90$ or 28).  Thus, information on Category 6 in TESICO will
accumulate more rapidly than that on the other categories; similarly,
information on hospitalization/death in A5401 will accrue more quickly
than information on subjects who remain alive and unhospitalized at
day 28.  Intuitively, basing an interim analysis on all observed
outcomes will naively overrepresent $Y=6$ and $Y=1$ and lead to
potentially biased inference on the treatment effect parameters.

To characterize this issue more precisely, if $C$ is the time from a
subject's entry into the study to the time of an interim analysis,
then $Y$ is known at the time of an interim analysis if $C >
T$. Otherwise, the ascertainment time $T$ is censored at $C$ and $Y$
is not observed.  Basing the analysis on all subjects with $C > T$
without taking appropriate account of the fact that $Y$ is not
available for those with $C \leq T$ leads to the bias noted above.
These considerations suggest that a valid interim analysis can be
obtained by using only the data from enrolled subjects followed for
the full, maximum follow-up period $\Tf$, i.e., for whom $C \geq \Tf$.
In studies like those above involving a continuous outcome or ordinal
categorical outcome, as in TESICO, the standard interim analysis is
based on the estimator to be used at the final analysis using only the
data on subjects with $C \geq \Tf$, as there is no apparent general
approach to ``adjusting'' for the censoring.  In the case of a binary
outcome as in A5401, the standard interim analysis does use the
information on censoring; e.g., if the treatment effect is the risk
ratio, the estimator is the ratio of the treatment-specific
Kaplan-Meier estimators for the probability of death or
hospitalization at $\Tf = 28$ days.

A goal of an interim analysis is to stop the trial as early as
possible if there is sufficiently strong evidence to do so.  It is
thus natural to consider whether or not it is possible to make more
efficient use of the available data at the time of an interim analysis
to enhance precision and thus the strength of the evidence for
stopping.  One step toward increasing efficiency of interim analyses
would be a general approach to accounting for censoring for any
outcome and treatment effect parameter to allow incorporation of
partial information; e.g., in TESICO, a subject who is at day 45 since
study entry and still in the hospital at the time of an analysis, so
for whom $C = 45 < \Tf = 90$, can end up only in Categories 3--6,
so have $Y=3, 4, 5$, or 6, at 90 days.  In addition, there may be
baseline covariates as well as intermediate measures of the outcome or
other post-treatment variables that could be exploited to increase
precision at an interim analysis.

In this article, we propose a general group sequential framework for
clinical trials with a possibly censored, time-lagged outcome, which
leads to practical strategies for interim monitoring.  Treatment
effect estimators are proposed via application of semiparametric
theory\cite{LuTsiatis,TsiatisBook}, which dictates how censoring can
be taken into account and baseline and time-dependent covariate
information can be exploited in a principled way to increase precision
and thus yield stronger evidence for early stopping.  Estimation of
the risk ratio via treatment-specific Kaplan-Meier estimators as
described above emerges as a simple special case, which can be
improved upon through incorporation of covariates.  We show that the
test statistics based on these estimators have an independent
increments structure\cite{Scharfstein}, which allows standard software
for constructing stopping boundaries\cite{Pocock,OBF,LanDeMets} to be used.  Two
interim monitoring strategies are discussed: an information-based
monitoring approach under which the trial will continue, with possibly
a larger sample size than originally planned, until the full, target
statistical information accrues; and a fixed-sample size approach
appropriate in settings where the planned sample size cannot be increased
due to resource and other constraints.  We focus on the common case of
two treatments; extension of the developments to more than two
treatments is possible\cite{TimeLag} and could be adapted to group
sequential methods for multi-arm trials\cite{Hellmich,Wason}.

In Section~\ref{s:model} we introduce the basic statistical framework
and assumptions, and we sketch the estimation approach and state the
independent increments property in Section~\ref{s:inference}.  In
Section~\ref{s:interim}, we describe practical implementation of the
resulting approach to interim monitoring.  We demonstrate the
performance of the methods in a series of simulation studies in
Section~\ref{s:sims}, and we present a case study exemplifying the use
of the methods for a simulated trial based on TESICO.  Technical
details and sketches of proofs of results are given in the
Appendix.

\section{Statistical framework}\label{s:model}

\subsection{General model}\label{s:genmod}

As in Section~\ref{s:intro}, denote the outcome by $Y$.  Let $A$
denote the treatment indicator, where $A = 0$ (1) corresponds to
control (active/experimental treatment), and $\pi = \pr(A=1)$ is the
probability of being assigned to active treatment; and let $X$ be a
vector of baseline covariates.  Treatment effects are often
characterized in terms of a model for (features of) the distribution
of $(Y,A)$ or $(Y,A,X)$, which involves parameters $(\alpha^T,\beta)$,
where $\beta$ is the scalar treatment effect parameter of interest and
$\alpha$ is a vector of nuisance parameters, and the model is
parameterized such that $\beta=0$ corresponds to the null hypothesis
of no treatment effect.

In the case of the first example in Section~\ref{s:intro} of
continuous $Y$, $\beta = E(Y | A=1) - E(Y|A=0)$; equivalently,
\begin{equation}
  E(Y | A=a) = \alpha + \beta a, \hspace{0.15in} a= 0, 1.
  \label{eq:meandiff}
  \end{equation}
  For an ordinal categorical outcome with $c$ categories, as in TESICO
  with $c=6$, the outcome can be represented as either a scalar random
  variable $Y$ taking values $1, \ldots, c$ or a random vector
  $Y = \{ I(Cat=1),\ldots, I(Cat=c-1)\}$, where $Cat$ takes values
  $1,\ldots,c$.  Using the first definition, the treatment effect can
  be defined through an assumed proportional odds model
\begin{equation}
\logit\{\pr( Y \leq j | A=a)\} = \alpha_j + \beta a, \hspace{0.15in}
j=1,\ldots,c-1, \hspace{0.15in} \alpha =
(\alpha_1,\ldots,\alpha_{c-1})^T, \hspace{0.15in} a= 0, 1,
\label{eq:propoddsmodel}
\end{equation}
where $\logit(p) = \log\{ p/(1-p)\}$, so that $\beta$ is the log odds
ratio of interest. If the conditional (on $X$) treatment effect is of interest, replace
(\ref{eq:propoddsmodel}) by
$\logit\{\pr( Y \leq j | X=x, A=a)\} = \alpha_j + \beta a + \xi^T x$,
where now $\alpha = (\alpha_1,\ldots,\alpha_{c-1}, \xi^T)^T$.  If
$Y$ is binary as in A5401 and the relative risk (risk ratio)
$\pr(Y=1|A=1)/\pr(Y=1|A=0)$ is the focus, taking
\begin{equation}
  E(Y | A=a) = \exp(\alpha + \beta a), \hspace{0.15in} a= 0, 1,
\label{eq:binary}
\end{equation}
corresponds to log relative risk $\beta$.

In general, estimators for the parameter of interest $\beta$ in
(\ref{eq:meandiff})-(\ref{eq:binary}) and other models based on the
data available at the final analysis, at which time $(Y, A, X)$ are
known for all $n$ participants, are obtained by solving, jointly in
$\alpha$ and $\beta$, appropriate estimating equations.  Namely, with
independent and identically distributed (iid) data $(Y_i,A_i,X_i)$,
$i=1,\ldots,n$, available, and $p$ equal to the dimension of
$(\alpha^T,\beta)^T$, $\hatalpha$ and $\hatbeta$ solve in $\alpha$ and
$\beta$ equations of the form
\begin{equation}
\sumin \calM(Y_i, A_i, X_i; \alpha,\beta) = 0,
\label{eq:fullesteq}
  \end{equation}
where $\calM(Y,A,X; \alpha,\beta)$ is a $p$-dimensional vector of
functions such that $E\{ \calM(Y,A,X; \alpha_0,\beta_0)\} = 0$, and
$\alpha_0$ and $\beta_0$ are the true values of $\alpha$ and $\beta$
under the assumption that the model is correctly specified.  For
example, under models (\ref{eq:meandiff}) and (\ref{eq:binary})
\begin{equation}
 \mbox{(a)  } \calM(Y, A, X; \alpha,\beta) = \left( \begin{array}{c} 1 \\  A \end{array}\right)
( Y - \alpha - \beta A)  \hspace{0.15in} \mbox{ and } \hspace{0.15in}
\mbox{(b)  }\calM(Y, A, X; \alpha,\beta) = \left( \begin{array}{c} 1 \\  A \end{array}\right)
\{ Y - \exp(\alpha + \beta A) \},
\label{eq:commonM}
\end{equation}
respectively.  Writing
$\overline{Y}_a = \sumin Y_i I(A_i=a)/\sumin I(A_i=a)$, $a = 0, 1$,
the treatment-specific sample means, the estimator obtained from
(\ref{eq:commonM})(a) is $\hatbeta = \overline{Y}_1-\overline{Y}_0$
and that from (\ref{eq:commonM})(b) is
$\hatbeta = \log(\overline{Y}_1/\overline{Y}_0)$, which is the estimator for
the relative risk used in A5401.    Under model (\ref{eq:propoddsmodel}), with
$\expit(u) = e^u/(1+e^u)$, 
\begin{equation}
  \calM(Y, A, X; \alpha,\beta)  = \calD(A) 
  \left( \begin{array}{c}
              I(Y \leq 1) -\expit(\alpha_1 + \beta A) \\
                                           \vdots \\
                I(Y \leq c-1) -\expit(\alpha_{c-1} + \beta A)  \end{array}
                                     \right),
\label{eq:propoddsM}
\end{equation}
where $\calD(A)$ is a $(c \times c-1)$ matrix of functions of $A$; the
ML estimator\cite{Agresti} takes
$\calD(A) = D^T(A; \alpha,\beta) V^{-1}(A; \alpha,\beta)$, where
$D(A; \alpha,\beta)$ is the $(c-1 \times c)$ gradient matrix of the
vector
$\{ I(Y \leq 1) -\expit(\alpha_1 + \beta A) ,\ldots, I(Y \leq c-1)
-\expit(\alpha_{c-1} + \beta A)\}^T$ in (\ref{eq:propoddsM}) with
respect to $\alpha_1,\ldots,\alpha_{c-1},\beta$, and
$V(A; \alpha,\beta)$ is its $(c-1 \times c-1)$ conditional covariance
matrix given $A$.


In general, given a particular model and estimating equations
defined by the corresponding function $\calM(Y, A, X; \alpha,\beta)$,         
let $\calG(\alpha,\beta)$ be the last row of the $(p \times p)$ matrix
\begin{equation}
- \left[E\left\{ \frac{\partial \calM(Y, A, X;
      \alpha,\beta)}{\partial\alpha^T \partial \beta}
  \right\}\right]^{-1},
\label{eq:pbypmatrix}
\end{equation}
where the matrix inside the expectation is
the $(p \times p)$ matrix of partial derivatives of the $p$ components of
$\calM(Y, A, X; \alpha,\beta)$ with respect to $(\alpha^T,\beta)^T$.
Then, with $\calG(\alpha_0,\beta_0)$ denoting this expression
evaluated at $\alpha_0, \beta_0$,
\begin{equation}
  m(Y, A, X; \alpha_0,\beta_0) = \calG(\alpha_0,\beta_0) \calM(Y,A,X; \alpha_0,\beta_0)
\label{eq:fulldatainflfunc}
  \end{equation}
  is referred to as the influence function of the corresponding
  estimator for $\beta$ and has mean zero.  From the theory of
  M-estimation\cite{StefanskiBoos} and semiparametric
  theory\cite{TsiatisBook}, it can be shown the estimator $\hatbeta$
  obtained by solving in $\beta$ the estimating equation
  \begin{equation}
\sumin m(Y, A, X; \hatalpha,\beta) =0,
\label{eq:fulldataesteqn}
    \end{equation}
    where $\hatalpha$ is any root-$n$ consistent estimator for
    $\alpha$, has influence function (\ref{eq:fulldatainflfunc}).
    Tsiatis et al.\cite{TimeLag} show this explicitly in the case of
    (\ref{eq:propoddsM}).  Such estimators are consistent for the true
    value $\beta_0$ and asymptotically normal, where the variance of
    the limiting normal distribution of $n^{1/2}(\hatbeta-\beta_0)$ is
    equal to
    $\mbox{var}\{ m(Y, A, X; \alpha_0,\beta_0)\} = E\{ m(Y, A, X;
    \alpha_0,\beta_0)^2 \}$, so that approximate (large sample)
    standard errors and test statistics are readily derived.

    From semiparametric theory\cite{TsiatisBook}, there is a
    one-to-one correspondence between influence functions and
    estimators.  Thus, if the form of influence functions in a
    specific model involving a parameter $\beta$ can be derived,
    estimating equations leading to estimators for $\beta$ can be
    developed.  As we demonstrate in Section~\ref{s:inference},
    influence functions corresponding to estimators for $\beta$ based
    on the data available at an interim analysis can be derived from
    the influence function (\ref{eq:fulldatainflfunc}), and the resulting
    estimators exploit baseline and time-dependent covariate
    information to gain precision.    

    \subsection{Data and assumptions}

    To characterize the data that would be available at an interim
    analysis, we first describe more fully the data that would be
    available at the final analysis if the trial were carried out to
    completion.  Subjects enter the trial in a staggered fashion;
    thus, if the trial starts at calendar time 0, denote by $E$ the
    calendar time at which a subject enters the trial.  As in
    Section~\ref{s:intro}, let $T$ denote the time lag in ascertaining
    the outcome $Y$; thus, $T$ is the time since entry at which $Y$ is
    determined, measured on the scale of subject time.  We assume that
    $Y$ can be determined with certainty by the maximum follow-up
    period $\Tf$ for any subject, so that $\pr(T \leq \Tf) = 1$.  In
    addition to baseline covariates $X$, time-dependent covariate
    information may be collected on each participant up to the time
    $Y$ is ascertained.  Denote by $L(u)$ the vector of such
    information at time $u$ following entry into the study, and let
    $\Lbar(u) = \{ L(s): 0 \leq s \leq u\}$ be the history of the
    time-dependent covariate information through time $u$.  Thus,
    $\Lbar(T)$ represents the covariate history for a subject for whom
    $Y$ is ascertained after time lag $T$.

    With these definitions, for a trial with planned total sample size
    $n$, the data available at the final analysis are iid
  \begin{equation}
    \{E_i, X_i, A_i, T_i, Y_i, \Lbar_i(T_i)\}, \hspace{0.15in} i=1,\ldots,n;
    \label{eq:fulldata}
    \end{equation}
    we refer to (\ref{eq:fulldata}) as the full data. As in
    Section~\ref{s:genmod}, estimation of $\beta$ at the final
    analysis is based only on the data on $Y$, $A$, and possibly $X$
    (in the case of conditional inference), and $E$, $T$, and
    $\Lbar(T)$ are not used, and we call 
    (\ref{eq:fulldatainflfunc}) a full data influence function.

      Now consider the data that would be available at an interim
      analysis at calendar time $t$ following the start of the trial
      at calendar time 0.  It proves convenient for the developments
      in Section~\ref{s:inference} to represent these data in terms of
      the full data (\ref{eq:fulldata}) that would be available at the
      final analysis were the trial to be carried out to completion.
      At $t$, data will be observed only for subjects for whom
      $E \leq t$.  For such subjects, define $C(t) = t-E$ to be the
      censoring time, i.e., the time from a participant's entry into
      the study to the time of the interim analysis.  If the time lag
      $T$ a subject would have in ascertaining the outcome is such
      that $T \leq C(t)$, then $Y$ would be available at $t$;
      otherwise, $Y$ would not yet be observed.  Accordingly, define
      $U(t) = \min\{ T, C(t) \}$ and $\Delta(t) = I\{ T \leq C(t) \}$,
      so that $Y$ is available at the time of the interim analysis
      only if $\Delta(t) = 1$.  With these definitions, the data
      available at an interim analysis at calendar time $t$ can be
      represented as iid
\begin{equation}
\calO{t}_i = I(E_i \leq t)\Big[\, E_i, X_i, A_i, U_i(t), \Delta_i(t), \Delta_i(t)
Y_i, \Lbar_i\{ U_i(t)\} \, \Big], \hspace{0.15in} i = 1,\ldots,n,
\label{eq:obsdatat}
\end{equation}
where then $n(t) = \sumin I(E_i \leq t)$ is the number of subjects of
the $n$ planned enrolled in the trial by calendar time $t$.

As noted in Section~\ref{s:intro}, an interim analysis that uses all
of the available data, including those from subjects for whom
$T < \Tf$, can naively overrepresent some values of the outcome over
others.  In terms of (\ref{eq:obsdatat}), the data on which this naive
analysis would be based involve only subjects $i$ who are enrolled and
whose outcome is available, i.e., for whom
$I\{E_i \leq t),\Delta_i(t) = 1\} = 1$.  In contrast, a valid analysis that
uses only the data from subjects enrolled for at least the full,
maximum follow-up period $\Tf$ involves subjects $i$ for whom
$I\{E_i \leq t, C_i(t) \geq \Tf\} = 1$.  In the next section, we
appeal to semiparametric theory as noted at the end of
Section~\ref{s:genmod} to deduce methods yielding valid inference on
$\beta$ based on the available data (\ref{eq:obsdatat}) that can
improve substantially on this analysis and thus lead to more efficient
interim analyses.
  
\section{Inference based on interim data} \label{s:inference}

\subsection{Treatment effect estimation}
\label{ss:theory}

We first present general estimating equations using the available 
data (\ref{eq:obsdatat}) at an interim analysis at time $t$ that yield
treatment effect estimators offering gains in precision
relative to the estimator based only on subjects for whom
$I\{E_i \leq t, C_i(t) \geq \Tf\} = 1$.  Letting ``$\independent$''
denote statistical independence, assume that $X \independent A$, which
is guaranteed by randomization.  Also assume that
    \begin{equation}
E \independent \{X, A, T, Y, \Lbar(T) \};
\label{eq:eindep}
      \end{equation}
      (\ref{eq:eindep}) implies that subjects enter according to a
      completely random process, which is reasonable in many trials.
      Because $C(t) = t-E$, (\ref{eq:eindep}) also implies
      that $C(t) \independent \{X, A, T, Y, \Lbar(T) \}$.  We discuss
      weakening these assumptions in Section~\ref{s:discuss}.  We also
      require that $\pr\{ C(t) > \Tf\} > 0$, so that there is positive
      probability of seeing subjects for whom the final outcome has
      been ascertained at an interim analysis at $t$ and so that 
      the first interim analysis must occur at least $\Tf$ time units
      after the start of the trial.

      We first summarize the theoretical underpinnings of the practical,
      more efficient interim monitoring approach we propose in
      Section~\ref{s:interim}.  Under the above assumptions, if
      $m(Y, A, X; \alpha_0, \beta_0)$ is the influence function of a
      given estimator for a treatment effect parameter $\beta$ in a
      model for $(Y, A, X)$ as in Section~\ref{s:genmod}, so based on
      the full data (\ref{eq:fulldata}), then semiparametric theory
      yields that influence functions for estimators for $\beta$ based
      on the available data $\calO{t}$ in (\ref{eq:obsdatat}) at an
      interim analysis at time $t$ are of the form
\begin{equation}
  \label{eq:obsinflfunc}
\begin{aligned}
\frac{I(E\leq t)}{\pr(E \leq t)} &\left( \frac{\Delta(t) m(Y, A, X;
  \alpha_0, \beta_0)}{\calK_t\{ U(t)\} } + \int^t_0 \frac{d\Mc{t}(u)
  \mu(m,u; \alpha_0,\beta_0)}{\calK_t(u) }   \right.\\
&-(A-\pi) f(X) + \left.\int^t_0 d\Mc{t}(u) \left[ h\{ u,X,A,\Lbar(u)\} -
\mu(h,u) \right] \right),
\end{aligned}
\end{equation}
where $f(X)$ is an arbitrary function of $X$; $h\{ u,X,A,\Lbar(u)\}$
is an arbitrary function of $u$, $X$, $A$, and $\Lbar(u)$; and  
$$\calK_t(u) = \pr\{ C(t) \geq u| E \leq t\}, \hspace{0.15in}
\mu(m,u; \alpha_0, \beta_0) = E\{ m(Y, A, X; \alpha_0, \beta_0) | T
\geq u\}, \hspace{0.15in} \mu(h,u) = E\big[ h\{ u,X,A,\Lbar(u)\}  | T
\geq u \big],$$
$$d\Mc{t}(u) = d\Nc{t}(u) - d\Lamc{t}(u) \Yt{t}(u), \hspace{0.15in}
\Nc{t}(u) = I\{ U(t) \leq u, \Delta(t) = 0\}, \hspace{0.15in}
\Yt{t}(u) = I\{ U(t) \geq u\}, \hspace{0.15in} \Lamc{t}(u) =
-\log\{\calK_t(u) \}.$$
Here, $\calK_t(u)$ is the survival distribution for the censoring
variable $C(t)$ at the time of the interim analysis, $\Nc{t}(u)$ and
$\Yt{t}(u)$ are the censoring counting process and at-risk process,
and $\Lamc{t}(u)$ is the cumulative hazard function for censoring.

Let $\hatK_t(u)$ be the Kaplan-Meier estimator for $\calK_t(u)$ using
the data $\{U_i(t), 1-\Delta_i(t)\}$ for $i$ such that $E_i \leq t$,
and define $\hatLamc{t}(u) = -\log\{ \hatK_t(u) \}$,
$d\hatMci{t}(u) = d\Nci{t}(u) - d\hatLamc{t}(u) \Yt{t}_i(u)$, and
$\hatpi_t = \sumin I(E_i \leq t) A_i /n(t)$, the proportion of enrolled
subjects at $t$ assigned to active treatment.  Then it can be shown
that estimating equations corresponding to the influence functions in
(\ref{eq:obsinflfunc}) based on the available data (\ref{eq:obsdatat})
yielding estimators for $\beta$ are of the form
\begin{equation}
\sumin I(E_i \leq  t) \left[ \frac{\Delta_i(t) m\{Y_i,A_i,X_i;
    \hatalpha(t), \beta\}}{\hatK_t\{ U_i(t)\} } - (A_i - \hatpi_t) f(X_i)
  + \int^t_0 d\hatMci{t}(u) h\{ u,X_i,A_i,\Lbar_i(u)\} \right] = 0,
\label{eq:obsesteqn}
\end{equation}
where $\hatalpha(t)$ is a consistent estimator for
$\alpha$ based on the available data at $t$.   For a specific model,
corresponding full data influence
function $m(Y, A, X; \alpha_0, \beta_0)$, and choice of the functions
$f(X)$ and $h\{ u,X,A,\Lbar(u)\}$, to be discussed momentarily, an
estimator for $\beta$ based on the data available at interim analysis
time $t$ is the solution to (\ref{eq:obsesteqn}).  

Taking $f(X) \equiv 0$ and $h\{ u,X,A,\Lbar(u)\} \equiv 0$ in (\ref{eq:obsesteqn}) yields the
estimating equation
\begin{equation}
\sumin I(E_i \leq  t) \left[ \frac{\Delta_i(t) m\{Y_i,A_i,X_i;
    \hatalpha(t), \beta\}}{\hatK_t\{ U_i(t)\} }  \right]= 0, 
 \label{eq:ipwesteqn}
\end{equation}
whose solution is a so-called inverse probability weighted complete case (IPWCC)
estimator, which effectively bases estimation of $\beta$ on only
subjects for whom $Y$ is available at $t$, but with inverse weighting
by the censoring distribution ``adjusting'' appropriately for the lag
time in ascertaining the outcome.  Judicious nonzero choices of $f(X)$
and $h\{ u,X,A,\Lbar(u)\}$ facilitate exploiting baseline and
time-dependent covariate information to gain efficiency over the IPWCC
estimator solving (\ref{eq:ipwesteqn}) through the two rightmost
``augmentation'' terms in the bracketed expression in
(\ref{eq:obsesteqn}), leading to what is referred to as an augmented
inverse probability weighted complete case (AIPWCC) estimator for
$\beta$; the optimal such choices are discussed below.

A counterintuitive result from semiparametric theory is that, for any
arbitrary $f(X)$ and $h\{ u,X,A,\Lbar(u)\}$, it is possible to improve
the precision of the above estimators by replacing the Kaplan-Meier
estimator $\hatK_t(u)$ by treatment-specific Kaplan-Meier estimators
$\hatK_t(u, a)$, say, obtained using the data
$\{U_i(t), 1-\Delta_i(t)\}$ for $i$ such that $E_i \leq t$ and
$A_i=a$, $a=0,1$, even though because of (\ref{eq:eindep}) the
distribution of $C(t)$ is not treatment dependent.   This
substitution leads to influence functions for estimators for $\beta$
based on the available data of the form
\begin{equation}
  \label{eq:obsinflfunc2}
\begin{aligned}
\frac{I(E\leq t)}{\pr(E \leq t)} &\left( \frac{\Delta(t) m(Y, A, X;
  \alpha_0, \beta_0)}{\calK_t\{ U(t),A\} } + \int^t_0 \frac{d\Mc{t}(u,A)
  \mu(m,u, A; \alpha_0,\beta_0)}{\calK_t(u,A) }   \right.\\
&-(A-\pi) f(X) + \left.\int^t_0 d\Mc{t}(u,A) \left[ h\{ u,X,A,\Lbar(u)\} -
\mu(h,u,A) \right] \right),
\end{aligned}
\end{equation}
where now 
$$\calK_t(u,A) = \pr\{ C(t) \geq u| E \leq t, A\}, \hspace{0.1in}
\mu(m,u,A; \alpha_0, \beta_0) = E\{ m(Y, A, X\alpha_0, \beta_0) | T
\geq u,A\}, \hspace{0.1in} \mu(h,u,A) = E\big[ h\{ u,X,A,\Lbar(u)\}  | T
\geq u,A \big],$$
$$d\Mc{t}(u,A) = d\Nc{t}(u) - d\Lamc{t}(u,A) \Yt{t}(u), \hspace{0.15in}
\Lamc{t}(u,A) = -\log\{\calK_t(u,A) \}.$$
Estimating equations corresponding to
(\ref{eq:obsinflfunc2}) are then
\begin{equation}
\sumin I(E_i \leq  t) \left[ \frac{\Delta_i(t) m\{Y_i,A_i,X_i;
    \hatalpha(t), \beta\} }{\hatK_t\{ U_i(t), A_i\} } - (A_i - \hatpi_t) f(X_i)
  + \int^t_0 d\hatMci{t}(u, A_i) h\{ u,X_i,A_i,\Lbar_i(u)\} \right] = 0,
\label{eq:obsesteqn2}
\end{equation}
where now $\hatLamc{t}(u,a) = -\log\{ \hatK_t(u,a) \}$; and
$d\hatMci{t}(u,a) = d\Nci{t}(u) - d\hatLamc{t}(u,a) \Yt{t}_i(u)$.  The
estimating equations (\ref{eq:obsesteqn2}) with $f(X) \equiv 0$ and
$h\{ u,X,A,\Lbar(u)\} \equiv 0$,
\begin{equation}
\sumin I(E_i \leq  t) \left[ \frac{\Delta_i(t) m\{Y_i,A_i,X_i;
    \hatalpha(t), \beta\} }{\hatK_t\{ U_i(t), A_i\} } \right]=0, 
\label{eq:ipwesteqn2}
\end{equation}
yield an IPWCC estimator, and, again, nonzero choices of $f(X)$ and
$h\{ u,X,A,\Lbar(u)\}$ lead to an AIPWCC estimator.

When $Y$ is a binary outcome, as in study A5401, it can be shown that
the IPWCC estimator $\hatbeta(t)$ solving (\ref{eq:ipwesteqn2}) is
algebraically identical to the logarithm of the ratio of
treatment-specific Kaplan-Meier estimators for the probability of
death or hospitalization at $\Tf $ days. Thus, as noted in
Section~\ref{s:intro}, the standard estimator for the risk ratio at an
interim analysis is a special case of the general formulation here.
Moreover, because this estimator is equivalent to an IPWCC estimator,
it should be possible to obtain more efficient inference on the risk
ratio at an interim analysis via an AIPWCC estimator.

Semiparametric theory provides the optimal choices of $f(X)$ and
$h\{ u,X,A,\Lbar(u)\}$ yielding the most precise AIPWCC estimator
solving either of (\ref{eq:obsesteqn}) or (\ref{eq:obsesteqn2}), given by
\begin{equation}
  \begin{aligned}
f^{opt}(X) &= E\{ m(Y, A, X; \alpha_0, \beta_0) | X, A=1) - E\{ m(Y, A,
X; \alpha_0, \beta_0) | X, A=0), \\
&h^{opt}\{ u,X,A,\Lbar(u)\} =\frac{ E\{ m(Y, A, X; \alpha_0, \beta_0) | T
\geq u,  X, A, \Lbar(u) \} }{ \calK_t(u) }.
\end{aligned}
\label{eq:optfandh}
\end{equation}
The conditional expectations in (\ref{eq:optfandh}) are not likely to
be known in practice.  We propose an approach to approximating
$f^{opt}(X)$ and $h^{opt}\{ u,X,A,\Lbar(u)\}$ in
Section~\ref{s:interim}.  We recommend estimating $\beta$ at an
interim analysis at time $t$ by $\hatbeta(t)$ solving an estimating
equation of the form (\ref{eq:obsesteqn2}) with the approximations for
$f^{opt}(X)$ and $h^{opt}\{ u,X,A,\Lbar(u)\}$ substituted.

From semiparametric theory, estimators solving estimating equations of
the form (\ref{eq:obsesteqn}) or (\ref{eq:obsesteqn2}) are consistent
for $\beta_0$ (for $n(t)$ and $n$ large) and asymptotically normal,
where, as at the end of Section~\ref{s:genmod}, the variance of the
large sample distribution of $\hatbeta(t)$ can be obtained from the
variance of the corresponding influence function.  Thus, the resulting
approximate standard errors $SE\{ \hatbeta(t) \}$ can be used to form
a Wald-type test statistic, $T(t) = \hatbeta(t)/SE\{ \hatbeta(t) \}$
appropriate for addressing the null hypothesis of no treatment effect,
$\mbox{H}_0\!\!:\beta_0 = 0$.

We conclude this section by noting an important implication of these
results.  In the case where the full data (\ref{eq:fulldata}) are
available, as would be the case at the conclusion of the trial if not
stopped early, the preceding developments lead to covariate-adjusted
estimators for $\beta$ based on the full data that have the potential
to yield increased efficiency over the usual full data analyses
outlined in Section~\ref{s:genmod}.  In particular, considering
(\ref{eq:obsesteqn2}), if $t_{end}$ is the calendar time at which the
trial concludes with the full data accrued and outcomes for all
subjects ascertained, $I(E_i \leq t_{end}) = 1$, $\Delta_i(t_{end}) = 1$,
$\calK_{t_{end}} \{U_i(t_{end}),A_i\} = 1$, and
$d\hatMci{t_{end}}(u, A_i) =0$, $i=1,\ldots,n$, and (\ref{eq:obsesteqn2})
becomes
\begin{equation}
\sumin \big[ m\{Y_i,A_i,X_i;  \hatalpha(t_{end}), \beta\} - (A_i - \hatpi)
f(X_i) \big] = 0,
\label{eq:fullcovadj}
\end{equation}
where $\hatpi = n^{-1} \sumin A_i$, with corresponding influence function
\begin{equation}
m(Y, A, X; \alpha_0,\beta_0) - (A-\pi) f(X).
\label{eq:fullcovadjinfl}
  \end{equation}
  As above, the optimal choice of $f(X)$ leading to the most precise
  estimator solving (\ref{eq:fullcovadj}) is that given in
  (\ref{eq:optfandh}).  The estimating equation (\ref{eq:fullcovadj})
  is of the form of those in Zhang et al.\cite{Zhang}.  Thus, the
  proposed approach leads naturally to estimators for a final analysis that exploit
  baseline covariate information to improve efficiency through the
  ``augmentation term'' $(A-\pi) f(X)$.  

\subsection{Interim analysis}

In practice, interim analyses will be carried out at times
$t_1< \cdots < t_K$, with the possibility of stopping the trial early,
e.g., for efficacy if evidence of a large treatment effect emerges at
an interim analysis. That is, focusing on efficacy, the trial
may be stopped at the first interim analysis time at which the
relevant test statistic exceeds some appropriate stopping boundary;
that is, if
$$| T(t_j) | \geq b_j, \hspace{0.15in} j=1,\ldots,K,$$
for a two-sided alternative or $T(t_j) \geq$ or $\leq b_j$,
$j=1,\ldots,K$, for a one-sided alternative, where $b_j$,
$j=1,\ldots,K$, are the stopping boundaries.  As is well-studied in
the group sequential testing literature, the stopping boundaries
$b_1,\ldots,b_K$ are
chosen to take into account multiple comparisons and ensure that the
resulting procedure preserves the desired overall type 1
error\cite{TsiatisIBM,Pocock,OBF,LanDeMets}.  Standard
methods\cite{Pocock,OBF,LanDeMets} for deriving stopping boundaries
are based on the premise that the sequentially-computed test
statistics $T(t_1),\ldots,T(t_K)$ have the so-called independent
increments structure\cite{Scharfstein,Kim}.

In the Appendix, we sketch an argument demonstrating
that, with the optimal choices of $f(X)$ and $h\{ u,X,A,\Lbar(u)\}$
given in (\ref{eq:optfandh}), the proposed test statistics, properly
normalized, have the independent increments structure.  Owing to this
property, the practical strategies for interim monitoring presented in
Section~\ref{s:interim} can be implemented using standard software for
computation of stopping boundaries; in the simulations in
Section~\ref{s:sims}, we use the R package ldbounds\cite{ldbounds}.  
    
\section{Practical implementation and interim monitoring
  strategies}\label{s:interim}

\subsection{Treatment effect estimation}
\label{ss:twostep}

Generalizing the approach in Tsiatis et al.\cite{TimeLag} in the
special case of a proportional odds model (\ref{eq:propoddsmodel}), we
propose estimation of $\beta$ at an interim analysis at time $t$ using
an AIPWCC estimator solving (\ref{eq:obsesteqn2}), which can be
obtained via a two-step algorithm.

Assume that the treatment effect $\beta$ of interest is defined within a
model for which, given full data at the end of the study, the estimator
for $\beta$ would be obtained jointly with that for $\alpha$ by
solving an estimating equation of the form in (\ref{eq:fullesteq})
for a particular estimating function $\calM(Y, A, X; \alpha,\beta)$.
Because the optimal choices $f^{opt}(X)$ and $h^{opt}\{
u,X,A,\Lbar(u)\}$ in (\ref{eq:optfandh}) are not known, we approximate
them by linear combinations of basis functions.  Letting
$f_0(X), f_1(X), \ldots, f_M(X)$ be functions of $X$ specified by the
analyst, with $f_0(X) \equiv 1$, approximate $f^{opt}(X)$ by
\begin{equation}
  \sum^M_{m=0} \psi_m f_m(X).
\label{eq:fbasis}
  \end{equation}
Similarly, specify basis functions of $\{u, X, \Lbar(u)\}$, 
$h_1\{ u,X,\Lbar(u)\}, \ldots, h_L\{ u,X,\Lbar(u)\}$, and approximate
$h^{opt}\{ u,X,a,\Lbar(u)\}$ by 
\begin{equation}
  \sum^L_{\ell = 1} \phi_{a,\ell} h_\ell\{ u,X,\Lbar(u)\},
  \hspace*{0.15in} a=0, 1.
\label{eq:hbasis}
  \end{equation}
  With suitably chosen basis functions, experience in other
  contexts\cite{LuTsiatis,TimeLag,Zhang} suggests that this approach
  can lead to AIPWCC estimators that achieve substantial efficiency
  gains over IPWCC estimators.

  The AIPWCC estimator for $\beta$ obtained by substituting
  (\ref{eq:fbasis}) and (\ref{eq:hbasis}) in (\ref{eq:obsesteqn2}) has
  influence function (\ref{eq:obsinflfunc2}) with these same
  substitutions.  Because from semiparametric theory the variance of
  the estimator depends on the variance of the influence function, as
  at the end of Section~\ref{s:genmod}, we find the coefficients
  $\psi_m$, $m=1,\ldots,M$, and $\phi_{a,\ell}$, $\ell = 1,\ldots, L$,
  $a = 0, 1$, that minimize this variance, which, from the form of
  (\ref{eq:obsinflfunc2}), is a least squares problem, as detailed below.
 
  With these considerations, the two-step algorithm is as follows.  At
  an interim analysis at time $t$:

\noindent
{\em Step 1.}  Estimate $\alpha$ and $\beta$ by solving jointly in
$\alpha$ and $\beta$ 
$$\sumin I(E_i \leq  t) \left[ \frac{\Delta_i(t) \calM(Y_i,A_i,X_i;
    \alpha,\beta)}{\hatK_t\{ U_i(t),A_i\} } \right]= 0$$ to obtain
$\hatalpha(t)$ and $\hatbeta^{init}(t)$; $\hatbeta^{init}(t)$ is an
IPWCC estimator solving (\ref{eq:ipwesteqn2}).  Then obtain an
estimator $\hatG\{\hatalpha(t),\hatbeta^{init}(t)\}$ for
$\calG(\alpha,\beta)$.  If the expectation in (\ref{eq:pbypmatrix}) is
analytically tractable, $\hatG\{\hatalpha(t),\hatbeta^{init}(t)\}$ is
the last row of (\ref{eq:pbypmatrix}) with $\hatalpha(t)$ and
$\hatbeta^{init}(t)$ substituted for $\alpha$ and $\beta$; if not,
take the estimator $\hatG\{\hatalpha(t),\hatbeta^{init}(t)\}$ to be
the last row of
$$-\left[n(t)^{-1} \sumin I(E_i \leq t) \frac{\Delta_i(t)}{\hatK_t\{
  U_i(t),A_i\} } \left.\frac{\partial \calM(Y_i, A_i,
      X_i;\alpha,\beta)}{\partial\alpha^T \partial \beta}
  \right|_{\alpha=\hatalpha(t), \beta =\hatbeta^{init}(t)}
\right]^{-1}.$$
For each subject $i$ for whom $E_i \leq t$, based on
(\ref{eq:fulldatainflfunc}), construct
$$m\{Y_i, A_i, X_i ; \hatalpha(t),\hatbeta^{init}(t)\} = 
\hatG\{\hatalpha(t),\hatbeta^{init}(t)\}\calM\{Y_i,A_i,X_i; \hatalpha(t),\hatbeta^{init}(t)\}.$$\\*[-0.1in]

\noindent {\em Step 2.}  Estimate the coefficients $\psi_m$,
$m=1,\ldots,M$, and $\phi_{a,\ell}$, $\ell = 1,\ldots,L$, $a = 0, 1$,
in the approximations (\ref{eq:fbasis}) and (\ref{eq:hbasis}) by
``least squares,'' as suggested above.  Namely, for each subject $i$
for whom $E_i \leq t$, define the ``dependent variable''
 $$\hatY_i(t) = \frac{\Delta_i(t) m\{Y_i, A_i, X_i ;
   \hatalpha(t),\hatbeta^{init}(t)\} }{\hatK_t\{ U_i(t),A_i\} } +
 \int_0^t \frac{d\hatMci{t}(u,A_i) \hatmu\{m,u,A_i; \hatalpha(t),
   \hatbeta^{init}(t)\} } { \hatK_t(u,A_i ) },$$ where, in the
 integrand in the second term of the above expression, for $A_i = a$,
 $a = 0, 1$,
   \begin{align*}
\frac{ \hatmu\{m,u,a; \hatalpha(t),   \hatbeta^{init}(t)\} }
   { \hatK_t(u, a ) } = &\left\{ \sum_{k=1}^n I(E_k \leq t)
                          \Yt{t}_k(u) I(A_k=a)\right\}^{-1}  \\
     &\times \sum^n_{k=1} \left\{\frac{I(E_k \leq t) \Delta_k(t) m\{Y_k, A_k, X_k ;
   \hatalpha(t),\hatbeta^{init}(t)\} }{\hatK_t\{ U_k(t),a\} }
     \Yt{t}_k(u) I(A_k=a)\right\}.
   \end{align*}
   Likewise, for each of these subjects and suitably chosen basis
   functions as discussed above,   define the $M + 1 + 2L$ ``covariates''
   $$(A_i-\hatpi_t) f_m(X_i), \hspace{0.15in} m=0,1,\ldots,M; $$
   $$I(A_i=0) \int^t_0 d\hatMci{t}(u,0) \big[ h_\ell\{u,X_i,\Lbar_i(u)\}
   -\hatmu(h_\ell, u, 0)\big], \hspace{0.15in} \ell = 1,\ldots,L, $$
   $$I(A_i=1)\int^t_0 d\hatMci{t}(u,1) \big[ h_\ell\{u,X_i,\Lbar_i(u)\}
   -\hatmu(h_\ell, u, 1)\big], \hspace{0.15in} \ell = 1,\ldots,L,$$
where, for $a=0, 1$  
$$\hatmu(h_\ell, u, a) = \left\{  \sum_{k=1}^n I(E_k \leq t)
  \Yt{t}_k(u) I(A_k=a)\right\}^{-1} \sum^n_{k=1} h_\ell\{ u, X_k,
\Lbar_k(u)\} \Yt{t}_k(u) I(A_k=a).$$ Then obtain estimators
$\hatpsi_m$, $m=0, 1,\ldots,M$, and $\hatphi_{a,\ell}$,
$\ell = 1,\ldots, L$, $a = 0, 1$, by linear regression of $\hatY_i(t)$
on the above covariates.  Based on this regression, obtain ``predicted
values'' for each subject $i$ for whom $E_i \leq t$ as
\begin{align*}
Pred_i  = (A_i - \hatpi_t) \sum^M_{m=0} \hatpsi_m f_m(X_i) + 
 \int_0^t  d\hatMci{t}(u,A_i) \sum^L_{\ell=1} \hatphi_{A_i,\ell}
  \big[ h_\ell\{u,X_i,\Lbar_i(u)\}  -\hatmu(h_\ell, u, A_i)\big].
\end{align*}
The estimator for $\beta$ is then obtained as the one-step update
\begin{equation}
\hatbeta(t) = \hatbeta^{init}(t) - n(t)^{-1} \sumin I(E_i \leq t) Pred_i,
\label{eq:onestepbeta}
\end{equation}
and an approximate standard error  for $\hatbeta(t)$ is
given by
\begin{equation}
SE\{ \hatbeta(t)\} = n(t)^{-1} \left[\sumin I(E_i \leq t) \{ \hatY_i(t) - Pred_i\}^2\right]^{1/2}.
\label{eq:SEonestepbeta}
\end{equation}
By an argument similar to that in
the Supplementary Material of Tsiatis et al.\cite{TimeLag}, the
estimator (\ref{eq:onestepbeta}) is asymptotically equivalent to an
AIPWCC estimator solving (\ref{eq:obsesteqn2}).

In some settings, scant time-dependent covariate information may be
available.  Here, a special case of the general AIPWCC formulation
that still attempts to gain efficiency from only baseline covariates
$X$ is to solve an estimating equation of the form in
(\ref{eq:obsesteqn2}) but with $f(X)$ as in (\ref{eq:optfandh}) and
$h\{ u,X,A,\Lbar(u)\} = 0$.  Implementation is as above, but with the
``dependent variable'' in Step 2 regressed only on the $M+1$
``covariates'' $(A_i-\hatpi_t) f_m(X_i)$, $m=0,1,\ldots,M$, to obtain
estimators $\hatpsi_m$, $m=0,1,\ldots,M$, and by redefining
$Pred_i = (A_i-\hatpi_t) \sum_{m=0}^M \hatpsi_m f_m(X_i)$ in the
one-step update (\ref{eq:onestepbeta}) and its associated standard
error (\ref{eq:SEonestepbeta}).  For definiteness, we refer to the
resulting estimator as ``AIPW1'' and that incorporating time-dependent
covariates above as ``AIPW2.''

\subsection{Interim analysis}
\label{ss:ess}

There is a vast literature on early stopping of clinical trials using
group sequential and other methods; such methods are readily applied
if the independent increments property holds.  We now discuss
information-based and fixed-sample size monitoring strategies for
using these approaches with the proposed treatment effect estimators,
for which, as argued in the Appendix and demonstrated
empirically in Section~\ref{s:sims}, the independent increments
property holds exactly or approximately.

In the general information-based monitoring approach\cite{TsiatisIBM},
monitoring and group sequential tests are based on the proportion of
the total information to be gained from the completed trial available
at interim analysis times $t$, where in the present context
information is approximated at time $t$ using the large-sample
approximate standard error of the relevant estimator $\hatbeta(t)$,
$SE\{ \hatbeta(t) \}$.  If a group sequential test is desired with
type 1 error $\alpha$ for testing $\mbox{H}_0\!\!: \beta_0 = 0$ and power
$(1-\gamma)$ against a clinically meaningful alternative value
$\beta_0 = \beta_A$, say, then the maximum information $MI$ required
to achieve this objective at the final analysis with a two-sided test
is
$$MI = \left(\frac{z_{\alpha/2} + z_\gamma}{\beta_A}\right)^2 IF,$$
where $z_\delta$ is the $(1-\delta)$ quantile of the standard normal
distribution, and $IF$ is an inflation factor to account for the loss
of power that results due to repeated testing relative to doing a
single final analysis.  For example, the inflation factor associated
with using O'Brien-Fleming stopping boundaries\cite{OBF} is modest,
equal to about 1.03; see Tsiatis\cite{TsiatisIBM}.  Information at an
interim analysis at time $t$ is approximated as
$$\mbox{Inf}(t) = \big[ SE\{ \hatbeta(t)\}\big]^{-2}.$$
Thus, the proportion of information at interim analysis time $t$ is
approximated as 
\begin{equation}
p(t) = \frac{\mbox{Inf}(t)}{MI}.
\label{eq:propinfoIBM}
\end{equation}
Given the proportion of information (\ref{eq:propinfoIBM}) together with,
e.g., the Lan-DeMets spending function\cite{LanDeMets}, standard
software can be used to obtain stopping boundaries such
that the resulting group sequential testing procedure has the desired
operating characteristics.

Typically, in determining the overall sample size for a clinical trial
to achieve the desired power to detect a meaningful difference at a
given level of significance, the analyst must make assumptions on the
values of nuisance parameters.  If these assumptions are not correct,
the trial could be underpowered.  An oft-cited advantage of
information-based monitoring is that interim analyses would continue
until the information $\mbox{Inf}(t)$ achieves the maximum
information $MI$, guaranteeing the desired operating characteristics
regardless of the values of nuisance parameters.  However, if the
assumptions leading to the target sample size are not correct, the
information available at the time this planned sample size is reached
and all participants have the outcome ascertained may be less than
$MI$.  If evidence emerges during the trial that $MI$ is unlikely to
be achieved, the sample size might be reestimated and increased so
that the full information threshold $MI$ is met.

In many trials, however, resource constraints or other factors may
make exceeding the originally planned sample size impossible, thus
rendering principled information-based monitoring infeasible.  If a
fixed, maximum sample size $n_{max}$, say, is planned and inalterable,
then the proportion of information available at an interim analysis at
any time $t$ on which stopping boundaries can be determined must
instead be based on $n_{max}$.  In terms of the data
(\ref{eq:obsdatat}) available at an interim analysis at $t$, as
before, $n(t) = \sumin I(E_i \leq t)$ is the number of subjects who
have enrolled by time $t$; of these subjects,
$n_A(t) = \sumin I\{E_i \leq t, C_i(t) \geq \Tf\}$ is the number who
have been enrolled for the maximum follow-up period and thus have the
outcome ascertained with certainty, and it is likely that
$n_A(t) < n(t)$.  In general, a typical interim analysis at $t$ for
fixed $n_{max}$ would be based only on the data from these $n_A(t)$
subjects, and, accordingly, the proportion of information available at
$t$ would be $p(t) = n_A(t)/n_{max}$, with $p(t) = 1$ at the final
analysis.  However, here, the proposed IPWCC and AIPWCC estimators
allow censoring due to the time lag in ascertaining the outcome to be
taken into account and incorporation of covariates to increase
efficiency, so make use of additional information in
(\ref{eq:obsdatat}) beyond that available on just the $n_A(t)$
subjects for whom the outcome has been ascertained by time $t$.  Thus,
if monitoring is based on test statistics constructed from these
estimators, the proportion of information available at $t$ should be
between $n_A(t)/n_{max}$ and $n(t)/n_{max}$.

With these considerations, for fixed-sample size monitoring, we
propose characterizing the proportion of information available at an
interim analysis at time $t$ in terms of what we refer to as the
effective sample size $n_{ESS}(t)$, say, at $t$.  Intuitively, we
define $n_{ESS}(t)$ to be the number of participants, had they been
enrolled for the maximum follow-up period $\Tf$ and had their outcome
ascertained with certainty, that would be required to lead to
an estimator for $\beta$ based only on data from such subjects with 
the same precision as that achieved by an IPWCC or AIPWCC estimator
for $\beta$ based on all of the available data at $t$.  The proportion
of information at $t$ would then be $n_{ESS}(t)/n_{max}$.

To define effective sample size formally, with $n^*$ subjects for whom
the outcome has been fully ascertained, indexed by $j = 1,\ldots,n^*$,
consider the estimator $\hatbeta^*$ obtained by solving in $\beta$ the
full data estimating equation (\ref{eq:fulldataesteqn}) based on these
$n^*$ subjects,
$$\sum_{j=1}^{n^*} m(Y_j, A_j, X_j; \hatalpha, \beta) = 0$$
for some consistent estimator $\hatalpha$.  Then from the
semiparametric theory, $\hatbeta^*$ has standard error approximately
equal to the square root of
$\mbox{var}\{ m(Y, A, X; \alpha_0, \beta_0) \}/n^*.$ The effective
sample size at an interim analysis at time $t$ for the IPWCC estimator
$\hatbeta(t)$ calculated using the available data (\ref{eq:obsdatat})
at $t$ is the value $n^*$ such that
$\mbox{var}\{ m(Y, A, X; \alpha_0, \beta_0) \}/n^* = SE\{\hatbeta(t)
\}^2$.  Accordingly, define the effective sample size when monitoring
is based on the IPWCC estimator as
\begin{equation}
n_{ESS}(t) = \frac{ \mbox{var}\{ m(Y, A, X; \alpha_0, \beta_0) }{  SE\{\hatbeta(t)  \}^2 }.
\label{eq:essipw}
\end{equation}
Because $\mbox{var}\{ m(Y, A, X; \alpha_0, \beta_0)\}$ is not known, in
practice we must estimate it based on the available data, which can be accomplished via the estimator
 \begin{equation}
\widehat{\mbox{var}}\{ m(Y, A, X; \alpha_0, \beta_0)\} = n(t)^{-1}
\sumin I(E_i \leq t) \frac{\Delta_i(t) m\{Y_i,A_i,X_i;
  \hatalpha(t),\hatbeta(t) \}^2 }{\hatK_t\{ U_i(t),A_i\} }.
\label{eq:varest}
   \end{equation}
   Thus, in practice, we obtain the approximate effective sample size as
\begin{equation}
   n_{ESS}(t) = \frac{ \widehat{\mbox{var}}\{ m(Y, A, X; \alpha_0,
     \beta_0)\} }{ SE\{\hatbeta(t) \}^2 }.
\label{eq:nessapprox}
\end{equation}   

The effective sample size for an AIPWCC estimator $\hatbeta(t)$
(either AIPW1 or AIPW2) calculated using the available data at $t$
via the two-step algorithm is defined similarly, but with the full
data influence function $m(Y, A, X; \alpha_0, \beta_0)$ in
(\ref{eq:essipw}) replaced by the influence function
$m(Y, A, X; \alpha_0,\beta_0) - (A-\pi) f^{opt}(X)$ as in
(\ref{eq:fullcovadjinfl}).  An estimator for 
$\mbox{var}\{ m(Y, A, X; \alpha_0,\beta_0) - (A-\pi) f^{opt}(X) \}$
based on the available data is given by
\begin{equation}
\widehat{\mbox{var}}\{ m(Y, A, X; \alpha_0, \beta_0)\} = n(t)^{-1}
\sumin I(E_i \leq t) \frac{\Delta_i(t) [ m\{Y_i,A_i,X_i;
  \hatalpha(t),\hatbeta(t) \} - Pred^*_i ]^2 }{\hatK_t\{ U_i(t),A_i\} },
\hspace{0.15in}
Pred^*_i =  (A_i-\hatpi_t) \sum_{m=0}^M \hatpsi_m  f_m(X_i), 
\label{eq:varestaipw}
\end{equation}
where now the ``predicted values'' $Pred^*_i$ are obtained by a
weighted least squares regression with ``dependent variable''
$m\{Y, A, X; \hatalpha(t), \hatbeta(t)\}$, ``covariates''
$(A_i-\hatpi_t) f_m(X_i)$, $m=0, 1, \ldots, M$, and ``weights''
$\Delta_i(t)/\hatK_t\{ U_i(t),A_i\}$.  Thus, for $\hatbeta(t)$ the
AIPW1 or AIPW2 estimator, $n_{ESS}(t)$ is defined as in
(\ref{eq:nessapprox}) with the numerator given by
(\ref{eq:varestaipw}).

With the appropriate definition of $n_{ESS}(t)$, we approximate the
corresponding proportion of information available at $t$ with interim
analyses based on an IPWCC or AIPWCCC estimator as
    \begin{equation}
      p(t) = \frac{n_{ESS}(t)}{n_{max}}.
      \label{eq:propinfoESS}
    \end{equation}
   From (\ref{eq:varest}) and (\ref{eq:varestaipw}), $p(t) = 1$ at the
   final analysis.   As for information-based monitoring, given the proportion of
 information (\ref{eq:propinfoESS}), one can use standard software
 with, e.g., the Lan-DeMets spending function\cite{LanDeMets} to obtain
 stopping boundaries.

 In the simulation studies in the next section, we study the methods
 under fixed-sample monitoring, as in our experience this approach is
 most common in practice.  Moreover, while performance of
 information-based monitoring with statistics that possess the
 independent increments property has been
 well-studied\cite{Scharfstein,Mehta}, because our approach to
 characterizing proportion of information in fixed-sample monitoring
 based on the proposed effective sample size measure is new,
 evaluation of its performance is required.

\section{Simulation studies}\label{s:sims}


We present results from several simulation studies, each involving
10000 Monte Carlo replications.  For each simulation scenario, we
considered a uniform enrollment process during the calendar time
interval $[0, E_{max}]$ with a maximum, fixed sample size $n_{max}$
and maximum follow up time $\Tf$, and fixed-sample size monitoring
with interim analyses planned at calendar times $t_1 < \cdots < t_K$
and a final analysis at time $t_{end} = E_{max}+\Tf = t_{K+1}>t_K$,
for a total of $K+1$ possible analyses.  For simplicity, in each
scenario, we took $n_{max}$ to be the sample size required to achieve
roughly 80\% or 90\% power for a single analysis at $t_{end}$ and did
not include an inflation factor\cite{TsiatisIBM}.  At each of the
$K+1$ analysis times, we estimated the relevant treatment effect
parameter $\beta$ four ways:
\begin{enumerate}[i]
\item using the estimator $\hatbeta_{\Tf}(t_j)$ obtained by carrying out
  the full data analysis based only on subjects enrolled for at least
  the maximum follow-up period $\Tf$, so using subjects $i$ for whom
  $I\{ E_i \leq t, C_i(t) \geq \Tf\} = 1$;

\item using the IPWCC estimator
  $\hatbeta_{IPW}(t_j) = \hatbeta^{init}(t_j)$ based on the available
  data (\ref{eq:obsdatat}), obtained at Step 1 of the two-step algorithm;

\item using the AIPWCC estimator $\hatbeta_{AIPW1}(t_j)$ based on the
  available data (\ref{eq:obsdatat}), obtained at Step 2 of the two-step algorithm using
  only baseline covariates $X$ to gain efficiency, as at the end of
  Section~\ref{ss:twostep};

\item using the AIPWCC estimator $\hatbeta_{AIPW2}(t_j)$ based on the
  available data (\ref{eq:obsdatat}) , obtained at Step 2 of the
  algorithm using both baseline and time-dependent covariates $X$ and $\Lbar(u)$.
  \end{enumerate}
  At the final analysis at time $t_{end}$ at which the outcome
  has been ascertained on all $n_{max}$ subjects,
  $\hatbeta_{\Tf}(t_{end})$ and $\hatbeta_{IPW}(t_{end})$ yield
  (versions of) the intended full data analysis; and
  $\hatbeta_{AIPW1}(t_{end})$ and $\hatbeta_{AIPW2}(t_{end})$ are
  identical and yield the covariate-adjusted analysis exploiting
  baseline covariate information discussed at the end of
  Section~\ref{ss:theory}.  In all scenarios, for the null
hypothesis $\mbox{H}_0\!\!: \beta_0 = 0$ and
  one-sided alternative hypotheses and level of significance
  $\alpha = 0.025$, we used the R package ldbounds\cite{ldbounds} with
  a Lan-DeMets spending function to compute both
  O'Brien-Fleming\cite{OBF} and Pocock\cite{Pocock} stopping
  boundaries at each analysis time $t_j$, $j=1,\ldots,K+1$.  For
  $\hatbeta_{\Tf}(t_j)$, this calculation was based on the proportion
  of information $p(t_j) = n_A(t_j)/n_{max}$; for each of the IPWCC
  and AIPWCC estimators $\hatbeta_{IPW}(t_j)$,
  $\hatbeta_{AIPW1}(t_j)$, and $\hatbeta_{AIPW2}(t_j)$, the stopping
  boundaries were obtained using the approximate proportion of
  information (\ref{eq:propinfoESS}) based on the relevant
  approximate effective sample size $n_{ESS}(t_j)$ given in (\ref{eq:nessapprox}).

  For each scenario, we present the following results from two
  simulation studies, one under $\mbox{H}_0$, so with data generated
  with $\beta_0=0$, and one under an alternative of interest $\beta_0=\beta_A$:
  \begin{enumerate}[i]
  \item for each estimator, Monte Carlo estimates of $\mbox{cov}\{
    \hatbeta(s), \hatbeta(t)\}$, $s < t$, 
    and $\var\{\hatbeta(t)\}$, $s, t \in \{t_1,\ldots,t_K,t_{end}\}$;
    if the independent increments property holds, 
    $\mbox{cov}\{ \hatbeta(s), \hatbeta(t)\} = \var\{ \hatbeta(t)\}$,
    $s < t$;

  \item for each estimator at each $t \in \{t_1,\ldots,t_K,t_{end}\}$,
    Monte Carlo mean and standard deviation of $\hatbeta(t)$, Monte
    Carlo mean of $SE\{ \hatbeta(t)\}$, and Monte Carlo mean square
    error (MSE) for $\hatbeta_{\Tf}(t)$ divided by that for $\hatbeta(t)$;

    \item for each estimator and stopping boundary, the Monte Carlo proportion of data sets
      for which $\mbox{H}_0$ was rejected, Monte Carlo estimate of expected
      sample size, and Monte Carlo estimate of expected stopping time.
      \end{enumerate}
      
      The first two simulation scenarios, demonstrating the methods
      for an ordinal categorical outcome and a binary outcome,
      respectively, are based on the TESICO study with $\Tf=90$ days,
      using the generative models adopted by Tsiatis et
      al.\cite{TimeLag}, with $n_{max}=602$, $E_{max} =240$ days, and
      $K=4$ interim analyses planned at calendar times
      $(t_1,\ldots,t_4) = (150, 195, 240, 285)$ days, with the final
      analysis at $t_{end} = 330$ days.  For each simulated subject,
      $A$ was generated as Bernoulli with $\pr(A=1) = \pi = 0.5$,
      where $a=0\,(1)$ corresponds to placebo (active agent).  To
      produce data for which the proportional odds model
      (\ref{eq:propoddsmodel}) holds, we generated
      $\Upsilon \sim U(0,1)$ and set
      $\Gamma= (1-A) \Upsilon + A \Upsilon (1/\mbox{OR})/\{1 -\Upsilon
      + \Upsilon (1/e^\beta)\}$, where as in (\ref{eq:propoddsmodel})
      $\beta$ is the log odds ratio, so that the distribution of
      $\Gamma$ given $A=1$ satisfies
      $\logit\{ \pr(\Gamma \leq u\,|\,A=1) \} = \logit\{ \pr(\Gamma
      \leq u \,|\, A=0) \} + \beta$.  For $Y$ an ordinal outcome, we
      took $\pr(Y = j | A=0) = 0.12, 0.23, 0.17, 010, 0.05, 0.33$ for
      $j=1,\ldots,6$ as in Table 1 of Tsiatis et al.\cite{TimeLag} and
      thus generated $Y$ according to in which interval $\Gamma$ fell
      as determined by the cutpoints
      $[0.00, 0.12, 0.35, 0.52, 0.62, 0.67, 1.00]$.  Then if
      $\Gamma < 0.52$, so $Y=1, 2,$ or 3, we took the time in hospital
      to be $H = \Tf\Gamma/0.52$ and the number of days at home and
      off oxygen as $\Tf - H$, and $T = \Tf$.  If
      $0.52 \leq \Gamma < 0.62$ or $0.62 \leq \Gamma < 0.67$,
      corresponding to $Y=4$ or 5, again $T = \Tf$; if
      $\Gamma \geq 0.67$, corresponding to death, time of death
     $T = (1-A)T_0 + A T_1$, where $T_0 \sim U(0,30)$ if $A=0$ and
      $T_1 \sim U(20,50)$ if $A=1$. A baseline covariate was generated
      as $X \sim \mathcal{N}\{ 1.5(\Upsilon -0.5), 1\}$, so that $X$
      is independent of $A$, correlated with $Y$, and does not affect
      the proportional odds model.  Two time dependent covariates were
      generated as $L_1(u) = I(\mathcal{W} < u)$, where
      $\mathcal{W} = H I(\Gamma<0.52) +\Tf I(\Gamma \geq 0.52)$, so
      that $L_1(u) = 1$ if the subject was still in the hospital at
      time $u$; and $L_2(u) = (\Tf-\mathcal{W}) L_1(u)$, the number of
      days the subject was expected to be out of the hospital at day
      $\Tf$, and $L(u) = \{L_1(u),L_2(u)\}$.  For a scenario with
      binary outcome, we generated the data according to the foregoing
      scheme, except that we defined $Y = 1$, corresponding to death,
      if $\Gamma \geq 0.67$, and $Y=0$ otherwise.

      For the first scenario with ordinal categorical outcome, we
      generated data as above under the null hypothesis, so with
      $\beta = \beta_0=0$, and $\beta=\beta_A = \log(1.5)$,
      corresponding to the alternative for which TESICO was powered
      (80\% at the final analysis with $n=602$)\cite{TimeLag}, and
      $\mbox{H}_A\!\!: \beta_0 > 0$.  Here, $\hatbeta_{\Tf}(t)$ is the
      ML estimator for $\beta$ in (\ref{eq:propoddsmodel}) obtained
      using the  R function polr in the MASS package\cite{MASS}.  Following
      Tsiatis et al.\cite{TimeLag}, to simplify implementation, we
      constructed $\hatbeta_{IPW}(t)$, $\hatbeta_{AIPW1}(t)$, and
      $\hatbeta_{AIPW2}(t)$ using the estimating function
      (\ref{eq:propoddsM}) with
      $\calD(A) = D^T(A; \alpha,\beta) V^{-1}_{ind}(A; \alpha,\beta)$,
      where $ V_{ind}(A; \alpha,\beta)$ is chosen according to the
      ``working independence'' assumption, so that with full data at
      the final analysis, $\hatbeta_{\Tf}(t_{end})$ and
      $\hatbeta_{IPW}(t_{end})$ are not identical.  As shown by
      Tsiatis et al.\cite{TimeLag} and borne out in the simulations
      below, the efficiency loss for $\hatbeta_{IPW}(t_{end})$
      relative to $\hatbeta_{\Tf}(t_{end})$ is negligible.

Under (a) the null hypothesis and
(b) the alternative, the
Monte Carlo sample covariance matrices of the 10000 estimates
$\{\hatbeta_{AIPW2}(t_1),\ldots,\hatbeta_{AIPW2}(t_4),\hatbeta_{AIPW2}(t_{end})\}$
are
\begin{equation}
\mbox{(a) }\left( \begin{array}{ccccc}
0.041 & 0.027 & 0.022 & 0.019 & 0.019\\
0.027 & 0.028 & 0.021 & 0.019 & 0.018\\
0.022 & 0.021 & 0.022 & 0.019 & 0.019\\
0.019 & 0.019 & 0.019 & 0.019 & 0.018\\
0.019 & 0.018 & 0.019 & 0.018 & 0.018 \end{array}\right),
\hspace{0.15in}
\mbox{(b) }\left( \begin{array}{ccccc}
0.042 & 0.027 & 0.022 & 0.019 & 0.019\\
0.027 & 0.029 & 0.022 & 0.019 & 0.019\\
0.022 & 0.022 & 0.022 & 0.019 & 0.019\\
0.019 & 0.019 & 0.019 & 0.019 & 0.019\\
0.019 & 0.019 & 0.019 & 0.019 & 0.019 \end{array}\right),
\label{eq:covmataipw}
\end{equation}                
clearly demonstrating that the independent increments property holds
approximately for this estimator.  Analogous results for the other
three estimators are given in the Appendix, showing that
the independent increments property holds approximately for all.

Under the null hypothesis and the alternative, Table~\ref{t:scenario1}
presents Monte Carlo mean and standard deviation, the Monte Carlo
average of standard errors $SE\{\hatbeta(t)\}$, and MSE
ratio defined above as the Monte Carlo MSE for the indicated
estimator divided by that for $\hatbeta_{\Tf}$.  Thus, the MSE ratio reflects 
efficiency of the indicated estimator relative to the usual ML
estimator using data only on subjects enrolled for the maximum
follow-up period $\Tf$.  From the table, all estimators are
consistent, with standard errors that track the Monte Carlo standard
deviations, under both hypotheses.  The efficiency gains over
$\hatbeta_{\Tf}(t)$ achieved at interim analyses by using any of
$\hatbeta_{IPW}(t)$, $\hatbeta_{AIPW1}(t)$, and $\hatbeta_{AIPW2}(t)$
are substantial.  The IPWCC estimator achieves gains solely through
accounting for censoring; the AIPWCC estimators improve on these gains
by additionally incorporating covariates.  Notably, $\hatbeta_{AIPW2}(t)$
yields a two-fold gain at the initial interim analysis.   For all
three estimators, the efficiency gains are most pronounced at the
early interim analyses where censoring is the most substantial and
diminish as censoring decreases as the trial progresses.  At the final
analysis, $\hatbeta_{\Tf}(t_{end})$ and $\hatbeta_{IPW}(t_{end})$ show
very similar performance, with $\hatbeta_{IPW}(t_{end})$ exhibiting
  minimal relative loss of efficiency, as noted above.  As expected,
$\hatbeta_{AIPW1}(t_{end})$ and $\hatbeta_{AIPW2}(t_{end})$ are
identical and, due to the incorporation of adjustment for baseline
covariates, result a 16\%-17\% gain in efficiency over the usual final analysis.

\begin{center}
\begin{table*}[t]%
  \caption{For Scenario 1 with ordered categorical outcome,
    performance of estimators for $\beta$ under (a) the null
    hypothesis $\beta=0$ and (b) the alternative
    $\beta = \log(1.5) = 0.405$ at each interim analysis time
    $(t_1,\ldots, t_4) = (150, 195, 240, 285)$ days and at the final
    analysis at $t_{end} = 330$ days MC Mean is the mean of 10000 Monte
    Carlo estimates; MC SD is the Monte Carlo standard deviation, Ave
    MC SE is the mean of Monte Carlo standard errors, and MSE ratio is
    the ratio of Monte Carlo mean square error for the AIPW2 estimator
    divided by that for the indicated estimator.\label{t:scenario1}}
  \centering
  \begin{tabular*}{500pt}{@{\extracolsep\fill}lrcccp{0.1in}rccc@{\extracolsep\fill}}
  \toprule
    & \textbf{MC Mean} & \textbf{MC SD}  & \textbf{Ave MC SE}  & \textbf{MSE ratio}
& &\textbf{MC Mean} & \textbf{MC SD}  & \textbf{Ave MC SE}  & \textbf{MSE ratio}\\
& \multicolumn{9}{@{}c@{}}{\textbf{(a) Null Hypothesis}} \\
  &\multicolumn{4}{@{}c@{}}{\textbf{$\hatbeta_{\Tf}(t)$}} &&\multicolumn{4}{@{}c@{}}{\textbf{ $\hatbeta_{IPW}(t)$}} \\
  \cmidrule{2-5}\cmidrule{7-10}
$t_1$ & -0.002 & 0.294  & 0.294  &  1.000 & & -0.004 &  0.232  &  0.232  &    1.603\\
$t_2$ & -0.002 & 0.221 &  0.221 &  1.000 & & -0.002 &  0.189  &  0.189 &     1.330\\
$t_3$ & -0.003 & 0.184 &  0.185 &  1.000& & -0.002 &  0.166 &  0.164  &    1.239\\
$t_4$ & -0.001 & 0.162 &  0.162 &   1.000& & -0.001 &  0.152 &  0.151 &     1.139\\
$t_{end}$ & 0.000 & 0.146  & 0.146 & 1.000 & & 0.000 &  0.147 &  0.146  &    0.991\\*[0.03in]
  &\multicolumn{4}{@{}c@{}}{\textbf{$\hatbeta_{AIPW1}(t)$}} &&\multicolumn{4}{@{}c@{}}{\textbf{$\hatbeta_{AIPW2}(t)$}} \\
  \cmidrule{2-5}\cmidrule{7-10}
$t_1$ & -0.004 & 0.221  & 0.221  &    1.775 && -0.005 & 0.203  & 0.198  &    2.095\\
$t_2$ & -0.002 & 0.178  & 0.178  &    1.534 && -0.002 &  0.168 &  0.165&  1.717\\
$t_3$ & -0.002 & 0.156  & 0.154  &    1.399 && -0.002 & 0.149  & 0.145 &     1.542\\
$t_4$ & -0.001 & 0.141  & 0 .140 &     1.327 && -0.001 &  0.138 &  0.136 &     1.380\\
$t_{end}$ & 0.000 & 0.135  & 0.135  &    1.169 && 0.000 & 0.135 &  0.135 &     1.169\\*[0.03in]
  \midrule
& \multicolumn{9}{@{}c@{}}{\textbf{(b) Alternative Hypothesis}} \\
  &\multicolumn{4}{@{}c@{}}{\textbf{$\hatbeta_{\Tf}(t)$}} &&\multicolumn{4}{@{}c@{}}{\textbf{ $\hatbeta_{IPW}(t)$}} \\
  \cmidrule{2-5}\cmidrule{6-9}
$t_1$ & 0.408 & 0.294  & 0.294  &    1.000 && 0.406 & 0.235  & 0.235 &     1.566\\
$t_2$ & 0.406 & 0.220  & 0.221  &    1.000 && 0.406 & 0.191  & 0.191 &     1.336\\
$t_3$ & 0.404 & 0.185  & 0.185  &    1.000 && 0.405 & 0.167  & 0.165 &    1.221\\
$t_4$ & 0.406 & 0.163  & 0.162  &    1.000 && 0.406 & 0.153  & 0.152  &    1.131\\
$t_{end}$ & 0.406 & 0.147  & 0.146  &    1.000 && 0.406 & 0.148  & 0.147  &    0.985\\*[0.03in]
 &\multicolumn{4}{@{}c@{}}{\textbf{$\hatbeta_{AIPW1}(t)$}} &&\multicolumn{4}{@{}c@{}}{\textbf{$\hatbeta_{AIPW2}(t)$}} \\
  \cmidrule{2-5}\cmidrule{6-9}
$t_1$ & 0.405 & 0.224  & 0.224  &    1.733 && 0.406 & 0.204 &  0.200 &    2.078\\
$t_2$ & 0.406 & 0.180  & 0.180  &    1.508 && 0.408 & 0.169  & 0.167 &    1.702\\
$t_3$ &  0.405 & 0.158  & 0.155  &    1.378 && 0.408 & 0.150  & 0.147  &   1.523\\
$t_4$ & 0.406 & 0.142  & 0.141  &    1.314&& 0.407 & 0.139  & 0.137   &   1.373\\
$t_{end}$ & 0.406 & 0.137  & 0.136   &   1.159 && 0.406 & 0.137  & 0.136 &     1.159\\
        \bottomrule
\end{tabular*}
\end{table*}
\end{center}

\begin{center}
\begin{table*}[ht]%
\caption{For Scenario 1 with ordered categorical outcome, interim
  analysis performance using each estimator with O'Brien-Fleming and
  Pocock stopping boundaries under (a) the null
    hypothesis $\beta=0$ and (b) the alternative
    $\beta = \log(1.5) = 0.405$, with maximum sample size
    $n_{max}=602$ and $t_{end} = 330$ days.  P(reject) is the
    proportion of Monte Carlo data sets for which the null hypothesis
    was rejected; MC E(SS) is the Monte Carlo average of number of
    subjects enrolled at the time the stopping boundary was crossed
    (standard deviation); and MC E(Stop) is the Monte Carlo
    average stopping time (days) (standard deviation).   The standard
    error for entries for P(reject) in (a) is $\approx 0.0016$.  
    \label{t:scenario1.stop}}
\centering
\begin{tabular*}{500pt}{@{\extracolsep\fill}lcccp{0.1in}ccc@{\extracolsep\fill}}
   \toprule
  & \textbf{P(reject)} & \textbf{MC E(SS)}  & \textbf{MC E(Stop)}
  &  & \textbf{P(reject)} & \textbf{MC E(SS)}  & \textbf{MC E(Stop)}\\
& \multicolumn{7}{@{}c@{}}{\textbf{(a) Null Hypothesis}} \\
  &\multicolumn{3}{@{}c@{}}{\textbf{O'Brien-Fleming}} & &\multicolumn{3}{@{}c@{}}{\textbf{Pocock}} \\
  \cmidrule{2-4} \cmidrule{6-8}
$\hatbeta_{\Tf}(t)$ & 0.024 & 601.9 (2.7)  &  329.3 (7.5) && 0.023 & 599.7 (21.4)  &   327.4  (19.5)\\
$\hatbeta_{IPW}(t)$ & 0.024 & 601.6 (7.8)  &   328.4 (12.1) && 0.024 & 598.8  (25.9)   &   326.7  (22.4)\\
$\hatbeta_{AIPW1}(t)$ & 0.024 & 601.7 (6.7)   &   328.5 (11.3) && 0.024 & 598.9  (25.5)  &   326.8  (22.0)\\
$\hatbeta_{AIPW2}(t)$ & 0.024 & 601.2 (11.4)  &   327.9  (14.9) && 0.027 & 598.0  (28.9)  &    326.1  (24.7)\\
  \midrule
  & \multicolumn{7}{@{}c@{}}{\textbf{(b) Alternative Hypothesis}} \\
  &\multicolumn{3}{@{}c@{}}{\textbf{O'Brien-Fleming}} & &\multicolumn{3}{@{}c@{}}{\textbf{Pocock}} \\
  \cmidrule{2-4} \cmidrule{6-8}
$\hatbeta_{\Tf}(t)$ & 0.784 & 592.7 (31.5)  &  284.2 (44.7) && 0.710 &  548.1 (85.3)  & 260.5 (68.7)\\
$\hatbeta_{IPW}(t)$ & 0.771 & 564.6 (62.5) &   257.1 (55.9) && 0.701 & 516.3 (100.3)  & 239.0 (74.7)\\
$\hatbeta_{AIPW1}(t)$ & 0.836 & 562.7 (62.7)  & 251.5 (53.4) && 0.774  & 508.3 (100.6) &  230.1 (72.2)\\
$\hatbeta_{AIPW2}(t)$ & 0.841 & 531.9 (81.7)  & 231.7 (58.0) && 0.783 & 483.4 (103.5) &   215.1 (72.3)\\
  \bottomrule
\end{tabular*}
\end{table*}
\end{center}  

Table~\ref{t:scenario1.stop} presents interim monitoring results using
each estimator with both O'Brien-Fleming and Pocock stopping
boundaries under the null hypothesis and under the alternative
$\beta = \log(1.5)$.  Under the null, the nominal level $\alpha =
0.025$ is achieved for all estimators.  Under the alternative, power
for $\hatbeta_{\Tf}(t)$ is slightly shy of the desired 80\%, as
expected with no inflation factor; by comparison, the AIPWCC
estimators yield improved power due to inclusion of covariate
information.   Under the alternative and both
types of boundaries, basing interim analyses on $\hatbeta_{IPW}(t)$,
$\hatbeta_{AIPW1}(t)$, and $\hatbeta_{AIPW2}(t)$ results in impressive
reductions in expected sample size and expected stopping time relative
to $\hatbeta_{\Tf}(t)$, with the gains especially impressive for
$\hatbeta_{AIPW2}(t)$.

For the second scenario with binary outcome, we generated data as
above under the null hypothesis and with the log odds ratio equal to
1.5, which implies a log relative risk (risk ratio) for death ($Y=1$)
of $\beta=\beta_0=\beta_A = \log(0.247/0.33) = -0.290$ as in (\ref{eq:binary})
and alternative hypothesis $\mbox{H}_A\!\!: \beta_0 < 0$.  We took
$n_{max} = 900$, which corresponds roughly to 90\% power to detect
this alternative.  The Monte Carlo sample covariance matrices of the
10000 estimates
$\{\hatbeta(t_1),\ldots,\hatbeta(t_4),\hatbeta(t_{end})\}$ for each of
the four estimators under both the null and alternative settings are
shown in the Appendix and exhibit patterns analogous to
those in (\ref{eq:covmataipw}), demonstrating that all estimators have
approximately the independent increments property.  Also shown in the
Appendix for each estimator at each analysis time are the
Monte Carlo mean and standard deviation, the Monte Carlo average of
standard errors $SE\{\hatbeta(t)\}$, and MSE ratio
defined above as the Monte Carlo MSE for the estimator divided by that
for $\hatbeta_{\Tf}(t)$ under the null and alternative hypotheses.  All
estimators are consistent, and standard errors are very close to the
Monte Carlo standard deviations.  Under both null and alternative
hypotheses, the estimator $\hatbeta_{IPW}(t)$, which takes censoring at
interim analysis times into account and as noted previously is
identical to the ratio of treatment-specific Kaplan-Meier estimators
often used in practice, achieves substantial efficiency gains over
$\hatbeta_{\Tf}(t)$, with a two-fold increase at the first interim
analysis and 24\% at the last at $t_4=285$ days.  These estimators are
equivalent, as expected, at the final analysis.  The AIPWCC estimators
$\hatbeta_{AIPW1}(t)$ and $\hatbeta_{AIPW2}(t)$ achieve even greater
gains.  Here, $\hatbeta_{AIPW2}(t)$ does not offer improved
performance over $\hatbeta_{AIPW1}(t)$; this behavior is not
surprising, as the time-dependent covariates
$L(u) = \{L_1(u),L_2(u)\}$ reflecting length of hospital stay do not
provide information on death.  As expected, these estimators are
identical at $t_{end}$ and offer 10\%-12\% gains in efficiency over
the standard analysis through adjustment for the baseline covariate.

Table~\ref{t:scenario2.stop} shows interim monitoring results using
each estimator with O'Brien-Fleming and Pocock stopping boundaries
under the null hypothesis and under the alternative
$\beta_A = -0.290$.  Again, overall testing procedures achieve the
nominal level.  Power gains over $\hatbeta_{\Tf}(t)$ under the
alternative are achieved using the AIPWCC estimators.  As for the
first scenario, basing interim analyses on $\hatbeta_{IPW}(t)$,
$\hatbeta_{AIPW1}(t)$, and $\hatbeta_{AIPW2}(t)$ yields substantial
reductions in expected sample size and stopping time over
$\hatbeta_{\Tf}(t)$ under the alternative, especially for the AIPWCC
estimators.

\begin{center}
\begin{table*}[t]%
\caption{For Scenario 2 with binary outcome, interim
  analysis performance using each estimator with O'Brien-Fleming and
  Pocock stopping boundaries under (a) the null
    hypothesis $\beta=0$ and (b) the alternative
    $\beta = \log(0.247/0.33) = -0.290$, with maximum sample size
    $n_{max}=900$ and $t_{end} = 330$ days.  Entries are as in
    Table~\ref{t:scenario1.stop}. The standard
    error for entries for P(reject) in (a) is $\approx 0.0016$. \label{t:scenario2.stop}}
\centering
\begin{tabular*}{500pt}{@{\extracolsep\fill}lcccp{0.1in}ccc@{\extracolsep\fill}}
   \toprule
  & \textbf{P(reject)} & \textbf{MC E(SS)}  & \textbf{MC E(Stop)}
  &  & \textbf{P(reject)} & \textbf{MC E(SS)}  & \textbf{MC E(Stop)}\\
& \multicolumn{7}{@{}c@{}}{\textbf{(a) Null Hypothesis}} \\
  &\multicolumn{3}{@{}c@{}}{\textbf{O'Brien-Fleming}} & &\multicolumn{3}{@{}c@{}}{\textbf{Pocock}} \\
  \cmidrule{2-4} \cmidrule{6-8}
$\hatbeta_{\Tf}(t)$ & 0.024 & 900.0 (2.3)  &  329.4 (6.5) && 0.022 & 896.3 (33.6)  &   327.4  (19.8)\\
$\hatbeta_{IPW}(t)$ & 0.024 & 898.8 (15.8)  &   328.0 (14.5) && 0.023 & 894.3  (42.3)   &   326.5  (23.7)\\
$\hatbeta_{AIPW1}(t)$ & 0.023 & 899.0 (14.0)   &   328.1 (13.8) && 0.025 & 894.1 (42.8)  &  326.3  (24.3)\\
$\hatbeta_{AIPW2}(t)$ & 0.024 & 899.0  (14.9)  &   328.0  (14.2) && 0.026 & 894.1 (42.7)  &  326.2  (24.4)\\
  \midrule
  & \multicolumn{7}{@{}c@{}}{\textbf{(b) Alternative Hypothesis}} \\
  &\multicolumn{3}{@{}c@{}}{\textbf{O'Brien-Fleming}} & &\multicolumn{3}{@{}c@{}}{\textbf{Pocock}} \\
  \cmidrule{2-4} \cmidrule{6-8}
$\hatbeta_{\Tf}(t)$ & 0.770 & 887.5 (44.3)  &  285.9 (44.0) && 0.690 & 827.2  (122.4) & 264.4 (67.2)\\
$\hatbeta_{IPW}(t)$ & 0.767 & 808.3 (121.4) &   241.3 (61.3) && 0.700 & 744.9 (157.6)  & 228.3 (76.7)\\
$\hatbeta_{AIPW1}(t)$ & 0.806 & 801.8 (122.5)  & 236.1 (59.6) && 0.746  & 733.2 (157.3) &  221.4 (74.7)\\
$\hatbeta_{AIPW2}(t)$ & 0.809 & 799.9 (123.4)  & 235.5 (59.7) && 0.748 & 731.6 (157.3) &   220.6 (74.7)\\
  \bottomrule
\end{tabular*}
\end{table*}
\end{center}  

The final simulation scenario involves a continuous outcome, with
$n_{max}=300$, $E_{max}=156$ weeks, and $\Tf=52$ weeks, so that
enrollment takes place over 3 years, with $K=4$ interim analyses
planned at calendar times 104, 130, 156, 182 weeks and the final
analysis at $t_{end} = 208$ weeks.  We generated treatment assignment
$A$ as Bernoulli with $\pr(A=1) = \pi = 0.5$, where $a=0\,(1)$
corresponds to placebo (active agent); and a categorical baseline
covariate $X_1$ was generated with $\pr(X_1=j) = 0.4, 0.3, 0.2, 0.1$
for $j=1,\ldots,4$.  With $(s_1,\ldots,s_5) = (0,4,12,24,52)$ weeks,
$\sigma=4.5$, $D$ a $(2 \times 2)$ matrix with
$\mbox{vech}(D) = (80,-0.5,0.08)$, and $\xi = (\xi_1,\xi_2)^T$, we
generated longitudinal measurements for each subject $i$ according to
the linear mixed effects model
$Z_{ij} = 65 I(X_1=1) + 60 I(X_1=2) + 55 I(X_1=3) + 49 I(X_1=4) +
\{\xi_1(1-A) + \xi_2 A\} s_j + b_{0i} +b_{1i} s_j + e_{ij}$, where
$b_i = (b_{0i},b_{1i})^T \sim \mathcal{N}(0, D)$ independent of
$e_{ij} \sim \mathcal{N}(0,\sigma^2)$.  The outcome for subject $i$ is
then $Y_i = Z_{i5}$, the longitudinal measure at $\Tf$ weeks.  As
would be likely in practice, we included in $X$ only the single
baseline covariate $Z_{i1}$, the value of the longitudinal measure at
time 0 and did not also include $X_1$, and we took the single
time-dependent covariate $L(u)$ at time $u$ to be the most recently
observed value of the longitudinal measurements $Z_{ij}$.  Under the
null hypothesis, $\xi = (-0.3, -0.3)^T$; under the alternative,
$\xi = (-0.3, -0.18)^T$ corresponding to
$\beta = \beta_0=\beta_A = 6.24$, for which $n_{max}=300$ yields
roughly 90\% power at the final analysis.  Results shown in the
Appendix demonstrate that the independent increments
property holds approximately for all estimators.  Here, the estimators
$\hatbeta_{\Tf}(t)$ and $\hatbeta_{IPW}(t)$ are identical because
$T = \Tf$ for all subjects, so that both are based only on subjects
followed for at least $\Tf$ weeks.  Standard errors
$SE\{ \hatbeta_{\Tf}(t)\}$ are obtained from the routine formula for a
difference in sample means assuming common treatment-specific
variance, while $SE\{ \hatbeta_{IPW}(t)\}$ follows from the IPWCC
influence function; these standard errors are asymptotically
equivalent but differ slightly for finite samples.  Incorporation of
the baseline covariate yields 10\%-20\% gains in efficiency; further
incorporation of the last outcome carried forward as a time-dependent
covariate leads to efficiency gains for $\hatbeta_{AIPW2}(t)$ of 34\%
to 47\%.

Interim monitoring results are shown in Table~\ref{t:scenario3.stop}
and are analogous to those in Tables~\ref{t:scenario1.stop} and
\ref{t:scenario2.stop}.  Under the null hypothesis, the Monte
Carlo rejection probability for $\hatbeta_{AIPW2}(t)$ with Pocock
boundaries exceeds slightly the nominal 0.025 level.  Again, under the
alternative, the AIPWCC estimators result in earlier expected sample
sizes and stopping times.

     \begin{center}
\begin{table*}[t]%
\caption{For Scenario 3 with continuous outcome, interim
  analysis performance using each estimator with O'Brien-Fleming and
  Pocock stopping boundaries under (a) the null
    hypothesis $\beta=0$ and (b) the alternative
    $\beta = 6.24$, with maximum sample size
    $n_{max}=300$ and $t_{end} = 208$ days.  Entries are as in
    Table~\ref{t:scenario1.stop}. The standard
    error for entries for P(reject) in (a) is $\approx 0.0016$. \label{t:scenario3.stop}}
\centering
\begin{tabular*}{500pt}{@{\extracolsep\fill}lcccp{0.1in}ccc@{\extracolsep\fill}}
   \toprule
  & \textbf{P(reject)} & \textbf{MC E(SS)}  & \textbf{MC E(Stop)}
  &  & \textbf{P(reject)} & \textbf{MC E(SS)}  & \textbf{MC E(Stop)}\\
& \multicolumn{7}{@{}c@{}}{\textbf{(a) Null Hypothesis}} \\
  &\multicolumn{3}{@{}c@{}}{\textbf{O'Brien-Fleming}} & &\multicolumn{3}{@{}c@{}}{\textbf{Pocock}} \\
  \cmidrule{2-4} \cmidrule{6-8}
$\hatbeta_{\Tf}(t)$ & 0.026 & 299.9 (3.1)  &  207.4 (5.5) && 0.025 & 298.6 (11.3)  &   206.2  (12.7)\\
$\hatbeta_{IPW}(t)$ & 0.026 & 299.8 (3.4)  &   207.4 (5.8) && 0.025 & 298.6  (11.5)   &   206.1  (12.9)\\
$\hatbeta_{AIPW1}(t)$ & 0.025 & 299.9 (2.6)   &   207.5 (5.0) && 0.026 & 298.6 (11.2)  &  206.2  (12.7)\\
$\hatbeta_{AIPW2}(t)$ & 0.026 & 299.8  (4.0)  &   207.0  (7.4) && 0.029 & 298.1 (13.2)  &  205.7  (14.5)\\
  \midrule
  & \multicolumn{7}{@{}c@{}}{\textbf{(b) Alternative Hypothesis}} \\
  &\multicolumn{3}{@{}c@{}}{\textbf{O'Brien-Fleming}} & &\multicolumn{3}{@{}c@{}}{\textbf{Pocock}} \\
  \cmidrule{2-4} \cmidrule{6-8}
$\hatbeta_{\Tf}(t)$ & 0.875 & 286.3 (26.0)  &  167.8 (30.6) && 0.823 & 259.9  (45.2) & 151.9 (41.9)\\
$\hatbeta_{IPW}(t)$ & 0.876 & 286.0 (26.4) &   167.4 (30.7) && 0.826 & 259.1 (45.4)  & 151.1 (41.9)\\
$\hatbeta_{AIPW1}(t)$ & 0.930 & 286.5 (25.5)  & 165.5 (28.9) && 0.896  & 255.9 (45.2) &  146.0 (39.6)\\
$\hatbeta_{AIPW2}(t)$ & 0.930 & 265.9 (37.3)  & 146.7 (31.2) && 0.892 & 239.8 (45.3) &   133.5 (37.8)\\
  \bottomrule
\end{tabular*}
\end{table*}
\end{center}  
 
We remark that all scenarios reflect the general result that basing
interim analyses on the proposed AIPWCC estimators leads to not only
more efficient inferences but also, because of the increased
precision, to a greater proportion of the total statistical
information being available at each interim analysis time than would
be available using the usual methods.  This feature implies that
O'Brien-Fleming boundaries will be less conservative for the proposed
estimators, leading to potential gains in expected sample size and
stopping times.

\section{Application}\label{s:example}

To demonstrate how use of the methods would proceed in practice as a
trial progresses, we consider the setting of TESICO with ordinal
categorical outcome, where the treatment effect of interest is the log
odds ratio $\beta$ in an assumed proportional odds model as in
(\ref{eq:propoddsmodel}). Because this trial is ongoing, we cannot
base this demonstration on data from the trial; accordingly, we
present use of the methods for a simulated data set generated according
to the first simulation scenario in Section~\ref{s:sims} with
$\beta = \log(1.5)$, which is based on this study. As in
Section~\ref{s:sims}, the planned maximum sample size is
$n_{max} = 602$, with full enrollment reached by $E_{max} = 240$ day.
Interim analyses are planned at 150, 195, 240, and 285 days, with the
final analysis to be conducted at $t_{end} = 330$ days, at which time
all $n_{max}$ participants will have completed the trial with their
outcomes ascertained. For definiteness, we use O'Brien-Fleming
stopping boundaries and focus on the null and alternative hypotheses
$\mbox{H}_0\!\!: \beta_0=0$ versus $\mbox{H}_A\!\!: \beta_0 > 0$, with
overall level of significance $\alpha = 0.025$.

Table~\ref{t:casestudy} shows how the trial would proceed if the
analyses were conducted at each interim analysis time $t$ using each
of the estimators $\hatbeta_{\Tf}(t)$, $\hatbeta_{IPW}(t)$,
$\hatbeta_{AIPW1}(t)$, and $\hatbeta_{AIPW2}(t)$. For each estimator,
the proportion of information at each of the interim analysis times
was calculated as described in Section~\ref{ss:ess} and was used to
obtain the stopping boundary. At the first interim analysis at 150
days, the proportion of information for the ML estimator
$\hatbeta_{\Tf}(t)$, which uses only those subjects among the $n(t)$
enrolled who have been followed for at least the maximum follow-up
time $\Tf$, is 0.257, whereas that for $\hatbeta_{AIPW2}(t)$ is 0.462,
almost twice as large. This striking difference is
reflected in the corresponding stopping boundaries: at the
first interim analysis at 150 days, the test statistic based on
$\hatbeta_{\Tf}(t)$ is 2.496, far from the boundary of 4.265, whereas
that based on $\hatbeta_{AIPW2}(t)$ is 2.966, almost reaching the
boundary of 3.099.  Basing the analyses on $\hatbeta_{AIPW2}(t)$
results in sufficient evidence to stop the trial at the second
interim analysis at 195 days with 487 subjects enrolled, while
sufficient evidence to stop using $\hatbeta_{\Tf}(t)$ does not emerge
until the fourth interim analysis at 285 days, with all $n_{max} =
602$ subjects enrolled.  Basing the analyses on $\hatbeta_{IPW}(t)$
and $\hatbeta_{AIPW1}(t)$ results in stopping the trial at 240 days,
with again all 602 planned subjects enrolled.

\begin{center}
\begin{table*}[t]%
  \caption{Interim analysis results for analyses at time
    $(t_1,\ldots,t_5) = (150, 195, 240, 285, 330)$ for the simulated
    TESICO trial; $n(t_j)$ is the number of subjects enrolled at
    $t_j$.  For each of the estimators $\hatbeta_{\Tf}(t)$ (the ML
    estimator based on data from all subjects followed for at least
    the maximum follow-up period $\Tf$ at $t$), the IPWCC estimator
    $\hatbeta_{IPW}(t)$, and the AIPWCC estimators
    $\hatbeta_{AIPW1}(t)$ and $\hatbeta_{AIPW2}(t)$, Est (SE) are the
    estimate $\hatbeta(t_j)$ (standard error $SE\{ \hatbeta(t_j)\}$)
    at $t_j$, $T$ is associated the Wald test statistic, $p(t_j)$ is
    the proportion of information at $t_j$, and $b_j$ is the
    O'Brien-Fleming stopping boundary.  Entries are boldfaced at the
    interim analysis at which the trial would be stopped using the
    indicated estimator. \label{t:casestudy}}
  \centering
  \begin{tabular*}{500pt}{@{\extracolsep\fill}ccccccp{0.07in}cccc@{\extracolsep\fill}}
      \toprule
&& \multicolumn{4}{@{}c@{}}{\textbf{$\hatbeta_{\Tf}(t)$}}   &&\multicolumn{4}{@{}c@{}}{\textbf{ $\hatbeta_{IPW}(t)$}} \\
       \cmidrule{3-6}\cmidrule{8-11} 
$t_j$ & $n(t_j)$ & Est (SE) & $T$ & $p(t_j)$ & $b_j$ &&  Est (SE) & $T$ & $p(t)$ &  $b_j$ \\
      \cmidrule{3-6}\cmidrule{8-11} 
150 & 368& 0.730 (0.292) & 2.496 & 0.257 & 4.265 && 0.547 (0.230) & 2.380 & 0.408 & 3.318  \\
195 & 487&0.619 (0.224) & 2.765 & 0.432 & 3.218 && 0.476 (0.193) & 2.473 &  0.581 & 2.733 \\
240 & 602&0.457 (0.187) & 2.445 & 0.611 & 2.657 && \textbf{0.423 (0.166)} & \textbf{2.551}  & \textbf{0.785} &\textbf{2.313} \\
285 & 602&\textbf{0.459 (0.162)} & \textbf{2.828} & \textbf{0.809} & \textbf{2.277} && --& -- & -- & -- \\
330 & 602&--  & -- &  -- & -- && -- & -- & -- & -- \\*[0.03in]
   & & \multicolumn{4}{@{}c@{}}{\textbf{ $\hatbeta_{AIPW1}(t)$}} && \multicolumn{4}{@{}c@{}}{\textbf{ $\hatbeta_{AIPW2}(t)$}} \\
       \cmidrule{3-6}\cmidrule{8-11} 
$t_j$ & $n(t_j)$ & Est (SE) & $T$ & $p(t)$ & $b_j$ &&  Est (SE) & $T$ & $p(t)$ &  $b_j$ \\
      \cmidrule{3-6}\cmidrule{8-11} 
150 & 368& 0.565 (0.218) & 2.586 & 0.382 & 3.444 && 0.590 (0.199) & 2.966 & 0.462 & 3.099 \\
195 & 487&0.497 (0.182) & 2.739 & 0.564 & 2.777 && \textbf{0.532 (0.167)} &\textbf{3.185} & \textbf{0.670}&  \textbf{2.521} \\
240 & 602& \textbf{0.409 (0.156)} & \textbf{2.615} & \textbf{0.757} & \textbf{2.362} && --& -- & -- & -- \\
285 & 602&--& -- & -- & -- && --& -- & -- & --  \\
330 & 602&-- & -- & -- & -- && -- & -- & -- & -- \\
\bottomrule
\end{tabular*}
\end{table*}
\end{center}  

\section{Discussion}\label{s:discuss}

We have proposed a general framework for design and conduct of group
sequential trials in the common situation where the outcome is 
known with certainty only after some time lag.  The methods account for 
censoring at the time of an interim analysis and incorporate baseline
and time-dependent evolving covariate information to improve
efficiency over standard analyses, facilitating earlier stopping with
potentially smaller numbers of enrolled subjects.  We have
demonstrated analytically and empirically that the proposed test
statistics possess the independent increments structure, so that
standard methods and software for specifying stopping boundaries can
be used.  The methods can be applied under both information-based
monitoring and fixed-sample monitoring strategies.  For the latter, we
have proposed the idea of effective sample size to characterize the
proportion of information available at an interim analysis.
Simulation studies demonstrate that the methods preserve the operating
characteristics of a monitored trial and that substantial reductions
in expected sample size and stopping time can be achieved.

As noted above, the proposed methodology is relevant in the large
class of problems where the outcome would be known with certainty for
all subjects at the final analysis.  For some trials with possibly
censored time-to-event outcome, interest may focus on the hazard ratio
under the assumption of proportional hazards.  Here, there is no
prespecified, maximum follow-up time $\Tf$ at which the outcome is
known with certainty, so that the proposed framework is not
applicable.  

The methods as presented are based on the assumption (\ref{eq:eindep})
that entry time is independent of all other variables, including
baseline covariates $X$, which implies that any interim analysis time
$t$ $C(t) \independent \{X, A, T, Y, \Lbar(T) \}$.   This assumption
is made tacitly in any clinical trial that focuses on inference on an
unconditional treatment effect parameter.  If the distribution of $X$
changes over the course of a trial, then (\ref{eq:eindep}) is
violated, and, intuitively, subjects enrolled at the time of an
interim analysis may not be representative of the population of
interest at the final analysis.  This is a general phenomenon and not
unique to our methodology.  If  (\ref{eq:eindep}) is violated in this
way, then the treatment effect parameter may not be static over time.
Under these circumstances, conditional (on $X$) inference may be more
appropriate; e.g., as in the case of the conditional proportional odds
model for ordinal categorical outcome in Section~\ref{s:genmod}.
Under the modified assumption $E \independent \{A, T, Y, \Lbar(T)
\} | X$ (independence conditional on $X$), the proposed methods can be
extended to support such conditional inference through incorporation
of relevant influence functions and modeling of the censoring
distribution as a function of $X$.


\section*{Acknowledgments}

The authors thank Drs. Birgit Grund and Michael Hughes for helpful discussions.

\section*{Supporting information}

R code implementing the simulations reported in Section~\ref{s:sims}
is available from the authors

\appendix

\renewcommand{\thesection}{\Alph{section}}

\setcounter{equation}{0}
\renewcommand{\theequation}{A\arabic{equation}}

\section{Demonstration of Independent Increments Property}

For definiteness, we consider AIPWCC estimators $\hatbeta(t)$ with
influence function (\ref{eq:obsinflfunc}) with $f^{opt}(X)$ and
$h^{opt}\{ u,X,A,\Lbar(u)\}$ in (\ref{eq:optfandh}) substituted for
$f(X)$ and $h\{u,X,A,\Lbar(u)\}$, i.e., the efficient influence
function; results for the IPWCC estimator follow from the argument.

We wish to describe the joint distribution of such estimators at
interim analysis times $t_1 < \cdots < t_K$ (calendar time).  It
suffices to consider the bivariate distribution of
$\{\hatbeta(s), \hatbeta(t)\}$, $s < t$.  The large-sample properties
of $\{\hatbeta(s), \hatbeta(t)\}$ are determined by the covariance
matrix of the corresponding influence functions for $\hatbeta(s)$ and
$\hatbeta(t)$.

Using 
\begin{equation}
\frac{I(E \leq t)}{\pr(E \leq t)} \frac{\Delta(t) m(Y,A,X;
  \alpha_0,\beta_0)}{\calK_t\{U(t)\}} = \frac{I(E \leq t)}{\pr(E \leq
  t)} \left[ m(Y,A,X;  \alpha_0,\beta_0) - \int^t_0 \frac{d\Mc{t}(u)}{\calK_t(u) } 
  \{m(Y,A,X; \alpha_0,\beta_0) - \mu(m,u; \alpha_0,\beta_0)\} \right],  \label{eq:equal1}
  \end{equation}
  \begin{align*}
    \mu(h^{opt},u) &= E\big[ h^{opt}\{u,X,A,\Lbar(u)\} | T \geq u\big] =
                       \frac{ E\left[ E\{ m(Y, A, X; \alpha_0, \beta_0) | T
                     \geq u,  X, A, \Lbar(u) \} | T \geq u \right] }{ \calK_t(u) } \\
    &= \frac{ E\{ m(Y, A, X; \alpha_0, \beta_0) | T \geq u \} }{
      \calK_t(u) } = \frac{ \mu(m,u; \alpha_0,\beta_0) }{\calK_t(u) },
  \end{align*}
and denoting the available data at interim analysis times $s$ and $t$
by $\calO{s}$ and $\calO{t}$ as in (\ref{eq:obsdatat}), the efficient
influence function for $\hatbeta(t)$ is given by
  \begin{align}
m_t(\calO{t}; \alpha_0,&\beta_0) = \frac{I(E \leq t)}{\pr(E \leq t)}
\left( \vphantom{\int^t_0 \frac{d\Mc{t}(u)}{\calK_t(u) }}\{m(Y,A,X;
  \alpha_0,\beta_0) - (A-\pi) f^{opt}(X)\} \right. \label{eq:first}\\
&- \left.\int^t_0 \frac{d\Mc{t}(u)}{\calK_t(u) } \big[ m(Y,A,X; \alpha_0,\beta_0)
- E\{ m(Y, A, X; \alpha_0, \beta_0) | T \geq u,  X, A, \Lbar(u) \}
\big] \right). \label{eq:second}
    \end{align}
Similarly, denote the efficient influence function for $\hatbeta(s)$ as
$m_s(\calO{s}; \alpha_0,\beta_0)$.  For brevity, we
suppress dependence of these influence functions on $\alpha_0,
\beta_0$ henceforth.

To demonstrate the independent increments property, it suffices to
show that
\begin{equation}
E\{ m_s(\calO{s}) m_t(\calO{t}) \} = \mbox{var}\{ m_t(\calO{t}) \} =
E\{ m_t(\calO{t})^2\}.
\label{eq:iiproperty}
\end{equation}
To this end, recalling that $C(t) = t-E$, it is convenient to define
$$\tilNc{t}(u) = I\{ C(t) \leq u\} = I(E \geq t-u), \hspace{0.15in}
\tilYt{t}(u) = I\{ C(t) \geq u\} = I(E \leq t-u),$$
and note that $\calK_t(u) = \pr\{ C(t) \geq u | E \leq t\} = \pr( E
\leq t-u | E \leq t)$.  Denoting the distribution of $E$ by $Q(u) =
\pr(E \leq u)$, with density $q(u)$, and recalling that $\Lamc{t}(u) =
-\log\{\calK_t(u) \}$, it follows that
$$\calK_t(u) = Q(t-u)/Q(t), \hspace{0.15in} d\Lamc{t}(u) =
\{q(t-u)/Q(t-u)\}\, du.$$
Thus, if we write 
\begin{equation}
  d\tilM(t-u) = d\tilNc{t}(u) -  d\Lamc{t}(u) \tilYt{t}(u) = -d I(E
  \geq t-u) - \frac{q(t-u)}{Q(t-u)} I(E \leq t-u) \, du,
  \label{eq:tilM}
\end{equation}
then using $\Nc{t}(u) = I\{ C(t) \leq u, T \geq u\}$ and $\Yt{t}(u) =
I\{ C(t) \geq u, T \geq u\}$, the martingale integral (\ref{eq:second}) can be written as
\begin{align}
&-\int^t_0 \frac{d \tilM(t-u)  I(T \geq u)}{Q(t-u)/Q(t)} \big[ m(Y,A,X)
                - E\{ m(Y, A, X) | T \geq u, X, A, \Lbar(u) \} \big] \nonumber \\
  &= -\int^t_0 \frac{d \tilM(x)  I(T \geq t-x)}{Q(x)/Q(t)} \big[ m(Y,A,X)
                - E\{ m(Y, A, X) | T \geq t-x, X, A, \Lbar(t-x) \} \big]
  \label{eq:changeofvar}
\end{align}
with change of variable $x = t-u$.  If we define the filtration
$\calF(x)$ to be the sigma algebra generated by
$\{ I(E \geq x), E I(E \geq x), Y, A, X, \Lbar(T)\}$, then the
integral (\ref{eq:changeofvar}) is the realization of a
$\calF(x)$-measurable martingale process.  Note that the filtration
$\calF(x)$ defines information about $E$ to the right (after) $x$
rather than as for the usual filtration that defines information to
the left (before) $x$.  Thus, (\ref{eq:second}) can be written as
\begin{align*}
-\frac{I(E \leq t)}{Q(t)}&\int^t_0 \frac{d \tilM(x)  I(T \geq t-x)}{Q(x)/Q(t)} \big[ m(Y,A,X)
                - E\{ m(Y, A, X) | T \geq t-x, X, A, \Lbar(t-x) \}  \big] \\
&= -\int^t_0 \frac{d \tilM(x)  I(T \geq t-x)}{Q(x)} \big[ \{m(Y,A,X)
                - E\{ m(Y, A, X) | T \geq t-x, X, A, \Lbar(t-x) \}  \big],
  \end{align*}
  and the variance of (\ref{eq:second}) is
  \begin{align*}
E&\left( \int^t_0 \frac{q(x)\, dx/Q(x) }{Q^2(x)} I(E \leq x) I(T \geq t-x)\big[ m(Y,A,X)
                - E\{ m(Y, A, X) | T \geq t-x, X, A, \Lbar(t-x) \}    \big]^2 \right)\\
&= \int^t_0 \frac{q(x)\, dx}{Q^2(x)} E\left( \big[ m(Y,A,X)
                - E\{ m(Y, A, X) | T \geq t-x, X, A, \Lbar(t-x) \}    \big]^2 I(T \geq t-x)\right).
    \end{align*}
    Consequently,
    \begin{equation}
      \label{eq:varmt}
\begin{aligned}      
\var\{ m_t(\calO{t}) \} &= \frac{\var\{ m(Y, A, X) - (A-\pi) f^{opt}(X)\}}{Q(t)}  \\
&+ \int^t_0 \frac{q(x)\, dx}{Q^2(x)} E\left( \big[ m(Y,A,X)
                - E\{ m(Y, A, X) | T \geq t-x, X, A, \Lbar(t-x) \}    \big]^2 I(T \geq t-x)\right),
\end{aligned}
\end{equation}      
and similarly for $\var\{ m_s(\calO{s})\}$.

From (\ref{eq:iiproperty}), we thus wish to show that $E\{
m_s(\calO{s} ) m_t(\calO{t} )\}$ is equal to $\var\{ m_t(\calO{t}) \}$ in
(\ref{eq:varmt}).  Using the preceding developments, we can write
\begin{align}
m_t(\calO{t}) &= \frac{I(E \leq t)}{Q(t)} \{ m(Y, A, X) - (A-\pi)
                f^{opt}(X)\} \label{eq:termone} \\
&- \int^t_0 \frac{d \tilM(x)}{Q(x)} \big[ \{m(Y,A,X)
                - E\{ m(Y, A, X) | T \geq t-x, X, A, \Lbar(t-x) \} \big] I(T \geq t-x)
\label{eq:termtwo}
\end{align}  
\begin{align}
m_s(\calO{s}) &= \frac{I(E \leq s)}{Q(s)} \{ m(Y, A, X) - (A-\pi)
                f^{opt}(X)\} \label{eq:termthree} \\
&- \int^s_0 \frac{d \tilM(x)}{Q(x)} \big[ \{m(Y,A,X)
                - E\{ m(Y, A, X) | T \geq s-x, X, A, \Lbar(s-x) \} \big] I(T \geq s-x);
\label{eq:termfour}
\end{align}  
accordingly,
$$E\{ m_s(\calO{s}) m_t(\calO{t})\} =
(\ref{eq:termone})\times(\ref{eq:termthree}) + (\ref{eq:termone}) \times (\ref{eq:termfour}) 
+ (\ref{eq:termtwo}) \times (\ref{eq:termthree}) +
(\ref{eq:termtwo}) \times (\ref{eq:termfour}).$$
We consider each of these terms in turn.  

Using $s \leq t$, it is straightforward that
\begin{align}
  E\{ (\ref{eq:termone}) \times (\ref{eq:termthree}) \} &= E\left[ \frac{I(E \leq s)}{Q(s)Q(t)} \{ m(Y, A, X) - (A-\pi)
                f^{opt}(X)\}^2\right] \nonumber \\
&= \frac{\var\{ m(Y, A, X) - (A-\pi) f^{opt}(X)\}}{Q(t)}. \label{eq:newtermone}
\end{align}
Similarly,
\begin{align}
  E\{ (\ref{eq:termone}) \times (\ref{eq:termfour}) \} &= E\left( \vphantom{\int^s_0 \frac{d \tilM(x)}{Q(x)} }
                                                 -\frac{I(E \leq s)}{Q(t)} \{ m(Y, A, X) - (A-\pi)
                f^{opt}(X)\} \right.\nonumber \\
 &\hspace*{0.2in}\times \left.\int^s_0 \frac{d \tilM(x)}{Q(x)} \big[ m(Y,A,X) - E\{ m(Y, A, X) | T \geq s-x, X, A, \Lbar(s-x) \}   \big] I(T \geq s-x)   \right) \nonumber \\
&= E\left( \vphantom{\int^s_0 \frac{d \tilM(x)}{Q(x)} }\frac{ \{ m(Y, A, X) - (A-\pi)   f^{opt}(X)\} }{Q(t)} \right.\nonumber \\
   &\hspace*{0.2in}\times \left.\int^s_0 \frac{d \tilM(x)}{Q(x)} \big[ m(Y,A,X)             - E\{ m(Y, A, X) | T \geq s-x, X, A, \Lbar(s-x) \}  \big] I(T \geq s-x) \right) \nonumber\\
&=0 \label{eq:newtermtwo}
\end{align}
because $\{ m(Y, A, X) - (A-\pi)   f^{opt}(X)\} $ is
$\calF(s)$-predictable, so that the expectation is zero by the
martingale property of stochastic integrals.  

By the martingale property,
\begin{equation}
  \label{eq:expect}
\begin{aligned}
E\{ (\ref{eq:termtwo})\times (\ref{eq:termfour}) \}  &= 
 \int^s_0 \frac{q(x) dx}{Q^2(x)} E \Big( \big[ m(Y,A,X) - E\{ m(Y, A, X) | T \geq t-x, X, A,   \Lbar(t-x) \}   \big]   \\
&\hspace*{0.2in}\times \big[ m(Y,A,X) - E\{ m(Y, A, X) | T \geq s-x, X, A, \Lbar(s-x) \}  \big] I(T \geq t-x)  \Big).
\end{aligned}
\end{equation}
The expectation in the intetgral in (\ref{eq:expect}) can be written
as 
\begin{align}
&E\left( \big[ m(Y,A,X) - E\{ m(Y, A, X) | T \geq t-x, X, A,  \Lbar(t-x) \} \big]^2 I(T \geq t-x) \right)
\nonumber\\
&-E\Big( \big[ E\{ m(Y, A, X) | T \geq t-x, X, A,  \Lbar(t-x) \} -E\{m(Y, A, X) | T \geq s-x, X, A,  \Lbar(s-x) \}
                         \big] \label{eq:middleterm}\\
&\times \big[ m(Y,A,X) - E\{ m(Y, A, X) | T \geq t-x, X, A,  \Lbar(t-x) \} \big]I(T \geq t-x) \Big).
                                                                 \nonumber
\end{align}
The difference of conditional expectations in brackets in (\ref{eq:middleterm}) is
some function $g\{ X, A, \Lbar(t-x)\}$, in which case the last two
lines of (\ref{eq:middleterm}) can be written as 
\begin{align*}
E&\Big( g\{ X, A, \Lbar(t-x)\} \big[ m(Y,A,X) - E\{ m(Y, A, X) | T \geq t-x, X, A,  \Lbar(t-x) \} \big]I(T \geq t-x) \Big)\\
&= E\left\{ E \left( g\{ X, A, \Lbar(t-x)\} \big[ m(Y,A,X) - E\{ m(Y,  A, X) | T \geq t-x, X, A,  \Lbar(t-x) \} \big]I(T \geq t-x) \right) 
\mid T \geq t-x, X, A,  \Lbar(t-x) \right\}\\
&=E\left( g\{ X, A, \Lbar(t-x)\} \big[  E\{ m(Y,  A, X) | T \geq t-x,     X, A,  \Lbar(t-x) \} -    E\{ m(Y,  A, X) | T   \geq t-x, X, A, 
                                         \Lbar(t-x) \} \big] I(T \geq
                                                t-x)\right) \\
  &= 0.
\end{align*}
It follows that 
\begin{equation}
E\{ (\ref{eq:termtwo}) \times (\ref{eq:termfour}) \}  = 
 \int^s_0 \frac{q(x) dx}{Q^2(x)} E\left( \big[ m(Y,A,X) - E\{ m(Y, A, X) | T \geq t-x, X, A,  \Lbar(t-x) \} \big]^2 I(T \geq t-x) \right).
\label{eq:newtermthree}
\end{equation}

Finally, consider
\begin{equation}
  \label{eq:finalterm}
\begin{aligned}
  E\{ (\ref{eq:termtwo}) \times (\ref{eq:termthree}) \} &= -E\left( \vphantom{\int^t_0\frac{d \tilM(x)}{Q(x)} }\frac{I(E \leq s)}{Q(s)} \{ m(Y, A, X) - (A-\pi)
    f^{opt}(X)\} \right. \\
  &\times \left.\int^t_0 \frac{d \tilM(x)}{Q(x)} \big[ \{m(Y,A,X) - E\{ m(Y, A, X) | T \geq t-x, X, A, \Lbar(t-x) \}  \big] I(T \geq t-x) \right).
\end{aligned}  
\end{equation}
We can write (\ref{eq:finalterm}) as
\begin{align}
-E&\left( \vphantom{\int^t_0\frac{d \tilM(x)}{Q(x)} }\frac{I(E \leq s)}{Q(s)} \{ m(Y, A, X) - (A-\pi)
    f^{opt}(X)\} \right. \nonumber \\
  &\times \left.\int^s_0 \frac{d \tilM(x)}{Q(x)} \big[ \{m(Y,A,X) - E\{ m(Y, A, X) | T \geq t-x, X, A, \Lbar(t-x) \}  \big] I(T \geq t-x) \right)
\label{eq:finalterma} \\
-E&\left( \vphantom{\int^t_0\frac{d \tilM(x)}{Q(x)} }\frac{I(E \leq s)}{Q(s)} \{ m(Y, A, X) - (A-\pi)
    f^{opt}(X)\} \right. \nonumber \\
  &\times \left.\int^t_s \frac{d \tilM(x)}{Q(x)} \big[ \{m(Y,A,X) - E\{ m(Y, A, X) | T \geq t-x, X, A, \Lbar(t-x) \}  \big] I(T \geq t-x) \right).
\label{eq:finaltermb}
\end{align}
Because $\{ m(Y, A, X) - (A-\pi)   f^{opt}(X)\} $ is
$\calF(s)$-predictable, (\ref{eq:finalterma}) is equal to zero.
Thus, consider (\ref{eq:finaltermb}).  Recalling from (\ref{eq:tilM})
that
$$d\tilM(x) = -d I(E \geq x) - \frac{q(x)}{Q(x)} I(E \leq x) dx$$
and noting that for $x \geq s$
$$I(E \leq s)\{-dI(E \geq x)\} = 0, \hspace{0.15in} I(E \leq s) I(E
\leq x) = I(E \leq s),$$
it follows that (\ref{eq:finaltermb}) can be written as
\begin{align}
E&\left( \int^t_s \frac{ \{ m(Y, A, X) - (A-\pi) f^{opt}(X)\}}{Q(s)}
  \frac{q(x) dx}{Q^2(x)} I(E \leq s) \big[ \{m(Y,A,X) - E\{ m(Y, A, X) | T \geq t-x, X, A, \Lbar(t-x) \}  \big] I(T \geq t-x) \right)\nonumber\\
&= \int^t_s \frac{q(x) dx}{Q^2(x)} E\left( \{ m(Y, A, X) - (A-\pi) f^{opt}(X)\} 
\big[ \{m(Y,A,X) - E\{ m(Y, A, X) | T \geq t-x, X, A, \Lbar(t-x) \}  \big] I(T \geq t-x) \right).\label{eq:nextterm}
\end{align}
Write the expectation in the integrand of (\ref{eq:nextterm}) as 
\begin{align}
E&\left( \big[ \{m(Y,A,X) - E\{ m(Y, A, X) | T \geq t-x, X, A, \Lbar(t-x) \}  \big]^2 I(T \geq t-x) \right)
\nonumber \\
&+E\left( \big[ E\{ m(Y, A, X) | T \geq t-x, X, A, \Lbar(t-x) \} - (A-\pi)
                          f^{opt}(X)\big] \right.\label{eq:finaltermd}\\
  &\times\left.\big[ \{m(Y,A,X) - E\{ m(Y, A, X) | T \geq t-x, X, A, \Lbar(t-x) \}  \big] I(T \geq t-x) \right).
\nonumber
\end{align}
Because the term in brackets in (\ref{eq:finaltermd}) is some function
$g^*\{ X, A, \Lbar(t-x)\}$, say, it follows by an argument similar to
that above that the last two lines are equal to zero, and thus
(\ref{eq:finalterm}) is equal to 
\begin{align}
E\{ (\ref{eq:termtwo})\times (\ref{eq:termthree}) \} = \int^t_s \frac{q(x) dx}{Q^2(x)} 
E\left( \big[ \{m(Y,A,X) - E\{ m(Y, A, X) | T \geq t-x, X, A, \Lbar(t-x) \}  \big]^2 I(T \geq t-x) \right).
\label{eq:newtermfour}
\end{align}

Combining the results in (\ref{eq:newtermone}), (\ref{eq:newtermtwo}),
(\ref{eq:newtermthree}), and (\ref{eq:newtermfour}) demonstrates the
desired result that
\begin{align*}
  E\{ m_s(\calO{s}) m_t(\calO{t})\}&= \frac{ \var\{ m(Y, A, X)  - (A-\pi) f^{opt}(X)\} }{Q(t)}\\
&+ \int^t_0 \frac{q(x) dx}{Q^2(x)} 
E\left( \big[ m(Y,A,X) - E\{ m(Y, A, X) | T \geq t-x, X, A, \Lbar(t-x) \}  \big]^2 I(T \geq t-x) \right),
  \end{align*}
which is (\ref{eq:varmt}).  

Note that the influence function for the IPWCC estimator solving
(\ref{eq:ipwesteqn}) is given by
(\ref{eq:obsinflfunc}) with $f(X) = h\{ u,X,A,\Lbar(u)\} \equiv 0$. Using
the equality (\ref{eq:equal1}) and the definition of $\tilM(x)$, the
influence function for the IPWCC estimator can be written as
\begin{align*}
m^{IPW}_t(\calO{t}) = \frac{ I(E \leq t)}{Q(t)} m(Y, A, X) - \int^t_0
  \frac{ d\tilM(x)}{Q(x)} \big[ m(Y,A,X) - E\{ m(Y, A, X) | T \geq t-x \}  \big]I(T \geq t-x).
  \end{align*}
That $E\{ m^{IPW}_s(\calO{s}) m^{IPW}_t(\calO{s}) \} = \var\{
m^{IPW}_t(\calO{t})\}$ follows by an argument analogous to that above,
demonstrating that the IPWCC estimator also has the independent
increments property.

In the practical implementation discussed in Section~\ref{s:interim},
the optimal choices $f^{opt}(X)$ and $h^{opt}\{u,X,A,\Lbar(u)\}$ are
approximated using linear combinations of basis functions.
Accordingly, the resulting AIPWCC estimators obtained via the two-step
algorithm may not be fully efficient and thus are not guaranteed to
have the independent increments property.  However, as demonstrated in
our simulation studies, because the approximations to $f^{opt}(X)$ and
$h^{opt}\{u,X,A,\Lbar(u)\}$ are often quite good, the estimators
themselves are good approximations to the efficient estimator and thus
exhibit behavior very close to that of independent increments, so that
the operating characteristics of the trial are preserved.

\renewcommand{\thesection}{\Alph{section}}

\setcounter{equation}{0}
\renewcommand{\theequation}{B\arabic{equation}}

\setcounter{table}{0}
\renewcommand{\thetable}{B\arabic{table}}

\section{Additional Simulation Results}

\noindent{\em Simulation Scenario 1: Ordinal Categorical Outcome:}
Under the null hypothesis, based on 10000 Monte Carlo data sets, the
Monte Carlo sample covariance matrices of
$\{\hatbeta(t_1),\ldots,\hatbeta(t_4),\hatbeta(t_{end})\}$, where
$\hatbeta(t)$ is each of $\hatbeta_{\Tf}(t)$, $\hatbeta_{IPW}(t)$,
$\hatbeta_{AIPW1}(t)$, and $\hatbeta_{AIPW2}(t)$, are given by
\begin{verbatim}
betahat_Tf
      [,1]  [,2]  [,3]  [,4]  [,5]
[1,] 0.086 0.049 0.034 0.026 0.021
[2,] 0.049 0.049 0.034 0.026 0.021
[3,] 0.034 0.034 0.034 0.026 0.021
[4,] 0.026 0.026 0.026 0.026 0.021
[5,] 0.021 0.021 0.021 0.021 0.021

betahat_IPW
[1,] 0.054 0.036 0.027 0.023 0.022
[2,] 0.036 0.036 0.027 0.023 0.022
[3,] 0.027 0.027 0.028 0.023 0.022
[4,] 0.023 0.023 0.023 0.023 0.022
[5,] 0.022 0.022 0.022 0.022 0.022

betahat_AIPW1
      [,1]  [,2]  [,3]  [,4]  [,5]
[1,] 0.049 0.031 0.023 0.019 0.018
[2,] 0.031 0.032 0.023 0.019 0.018
[3,] 0.023 0.023 0.024 0.020 0.019
[4,] 0.019 0.019 0.020 0.020 0.018
[5,] 0.018 0.018 0.019 0.018 0.018

betahat_AIPW2
      [,1]  [,2]  [,3]  [,4]  [,5]
[1,] 0.041 0.027 0.022 0.019 0.019
[2,] 0.027 0.028 0.021 0.019 0.018
[3,] 0.022 0.021 0.022 0.019 0.019
[4,] 0.019 0.019 0.019 0.019 0.018
[5,] 0.019 0.018 0.019 0.018 0.018
\end{verbatim}
Under the alternative $\beta_A = \log(1.5)$, the analogous Monte Carlo
sample covariance matrices are
\begin{verbatim}
betahat_Tf
      [,1]  [,2]  [,3]  [,4]  [,5]
[1,] 0.087 0.048 0.034 0.026 0.021
[2,] 0.048 0.049 0.034 0.026 0.021
[3,] 0.034 0.034 0.034 0.027 0.022
[4,] 0.026 0.026 0.027 0.027 0.022
[5,] 0.021 0.021 0.022 0.022 0.022

betahat_IPW
      [,1]  [,2]  [,3]  [,4]  [,5]
[1,] 0.055 0.036 0.028 0.023 0.022
[2,] 0.036 0.036 0.028 0.023 0.022
[3,] 0.028 0.028 0.028 0.024 0.022
[4,] 0.023 0.023 0.024 0.024 0.022
[5,] 0.022 0.022 0.022 0.022 0.022

betahat_AIPW1
     [,1]  [,2]  [,3]  [,4]  [,5]
[1,] 0.050 0.032 0.024 0.020 0.019
[2,] 0.032 0.032 0.024 0.020 0.019
[3,] 0.024 0.024 0.025 0.020 0.019
[4,] 0.020 0.020 0.020 0.020 0.019
[5,] 0.019 0.019 0.019 0.019 0.019

betahat_AIPW2
[1,] 0.042 0.027 0.022 0.019 0.019
[2,] 0.027 0.029 0.022 0.019 0.019
[3,] 0.022 0.022 0.022 0.019 0.019
[4,] 0.019 0.019 0.019 0.019 0.019
[5,] 0.019 0.019 0.019 0.019 0.019
\end{verbatim}
These results clearly demonstrate that the independent increments
property holds approximately for all estimators.

\vspace*{0.15in}

\noindent{\em Simulation Scenario 2:  Binary Outcome:}
Under the null hypothesis, based on 10000 Monte Carlo data sets, the
Monte Carlo sample covariance matrices of
$\{\hatbeta(t_1),\ldots,\hatbeta(t_4),\hatbeta(t_{end})\}$, where
$\hatbeta(t)$ is each of $\hatbeta_{\Tf}(t)$, $\hatbeta_{IPW}(t)$,
$\hatbeta_{AIPW1}(t)$, and $\hatbeta_{AIPW2}(t)$, are given by
\begin{verbatim}
betahat_Tf
      [,1]  [,2]  [,3]  [,4]  [,5]
[1,] 0.039 0.022 0.015 0.012 0.009
[2,] 0.022 0.021 0.015 0.011 0.009
[3,] 0.015 0.015 0.015 0.011 0.009
[4,] 0.012 0.011 0.011 0.011 0.009
[5,] 0.009 0.009 0.009 0.009 0.009

betahat_IPW
      [,1]  [,2]  [,3]  [,4]  [,5]
[1,] 0.018 0.013 0.011 0.009 0.009
[2,] 0.013 0.013 0.010 0.009 0.009
[3,] 0.011 0.010 0.010 0.009 0.009
[4,] 0.009 0.009 0.009 0.009 0.009
[5,] 0.009 0.009 0.009 0.009 0.009

betahat_AIPW1
      [,1]  [,2]  [,3]  [,4]  [,5]
[1,] 0.017 0.012 0.009 0.008 0.008
[2,] 0.012 0.012 0.009 0.008 0.008
[3,] 0.009 0.009 0.009 0.008 0.008
[4,] 0.008 0.008 0.008 0.008 0.008
[5,] 0.008 0.008 0.008 0.008 0.008

betahat_AIPW2
      [,1]  [,2]  [,3]  [,4]  [,5]
[1,] 0.017 0.012 0.009 0.008 0.008
[2,] 0.012 0.012 0.009 0.008 0.008
[3,] 0.009 0.009 0.009 0.008 0.008
[4,] 0.008 0.008 0.008 0.008 0.008
[5,] 0.008 0.008 0.008 0.008 0.008
\end{verbatim}
Under the alternative $\beta_A = \log(0.247/0.33)=-0.290$, the analogous Monte Carlo
sample covariance matrices are
\begin{verbatim}
betahat_Tf
      [,1]  [,2]  [,3]  [,4]  [,5]
[1,] 0.049 0.027 0.019 0.015 0.012
[2,] 0.027 0.027 0.019 0.014 0.012
[3,] 0.019 0.019 0.019 0.014 0.011
[4,] 0.015 0.014 0.014 0.014 0.011
[5,] 0.012 0.012 0.011 0.011 0.011

betahat_IPW
      [,1]  [,2]  [,3]  [,4]  [,5]
[1,] 0.023 0.017 0.013 0.012 0.012
[2,] 0.017 0.017 0.013 0.012 0.011
[3,] 0.013 0.013 0.013 0.011 0.011
[4,] 0.012 0.012 0.011 0.011 0.011
[5,] 0.012 0.011 0.011 0.011 0.011

betahat_AIPW1
      [,1]  [,2]  [,3] [,4] [,5]
[1,] 0.022 0.015 0.012 0.010 0.010
[2,] 0.015 0.015 0.012 0.010 0.010
[3,] 0.012 0.012 0.012 0.010 0.010
[4,] 0.010 0.010 0.010 0.010 0.010
[5,] 0.010 0.010 0.010 0.010 0.010
\end{verbatim}
\clearpage

\begin{verbatim}
betahat_AIPW2
      [,1]  [,2]  [,3] [,4] [,5]
[1,] 0.022 0.015 0.012 0.010 0.010
[2,] 0.015 0.015 0.012 0.010 0.010
[3,] 0.012 0.012 0.012 0.010 0.010
[4,] 0.010 0.010 0.010 0.010 0.010
[5,] 0.010 0.010 0.010 0.010 0.010
\end{verbatim}
These results clearly demonstrate that the independent increments
property holds approximately for all estimators.

Table~\ref{t:scenario2} presents performance of the estimators under
the null and alternative hypotheses.  
\begin{center}
\begin{table*}[t]%
  \caption{For Scenario 2 with binary outcome,
    performance of estimators for $\beta$ under (a) the null
    hypothesis $\beta=0$ and (b) the alternative
    $\beta = \log(0.247/0.33) = -0.290$ at each interim analysis time
    $(t_1,\ldots, t_4) = (150, 195, 240, 285)$ days and at the final
    analysis at $t_{end} = 330$ days MC Mean is the mean of 10000 Monte
    Carlo estimates; MC SD is the Monte Carlo standard deviation, Ave
    MC SE is the mean of Monte Carlo standard errors, and MSE ratio is
    the ratio of Monte Carlo mean square error for the AIPW2 estimator
    divided by that for the indicated estimator.\label{t:scenario2}}
  \centering
  \begin{tabular*}{500pt}{@{\extracolsep\fill}lrcccp{0.1in}rccc@{\extracolsep\fill}}
  \toprule
    & \textbf{MC Mean} & \textbf{MC SD}  & \textbf{Ave MC SE}  & \textbf{MSE ratio}
& &\textbf{MC Mean} & \textbf{MC SD}  & \textbf{Ave MC SE}  & \textbf{MSE ratio}\\
& \multicolumn{9}{@{}c@{}}{\textbf{(a) Null Hypothesis}} \\
  &\multicolumn{4}{@{}c@{}}{\textbf{$\hatbeta_{\Tf}(t)$}} &&\multicolumn{4}{@{}c@{}}{\textbf{ $\hatbeta_{IPW}(t)$}} \\
  \cmidrule{2-5}\cmidrule{7-10}
$t_1$ & 0.003 &  0.197  &  0.193  & 1.000&&  0.000 & 0.136  & 0.134  &    2.097\\
$t_2$ & 0.000 &  0.146  &  0.145   &    1.000&&  0.000 & 0.115  & 0.114   &   1.600\\
$t_3$ &  0.000 &  0.122  &  0.121   &     1.000&& -0.001 & 0.102  & 0.101  &    1.433\\
$t_4$ &  0.000 &  0.106  &  0.106    &    1.000&&  0.000 &  0096  & 0.095  &    1.237\\
$t_{end}$ & 0.000 &  0.096  &  0.095   &     1.000&& 0.000 &  0.096  & 0.095  &    1.000    \\*[0.03in]
  &\multicolumn{4}{@{}c@{}}{\textbf{$\hatbeta_{AIPW1}(t)$}} &&\multicolumn{4}{@{}c@{}}{\textbf{$\hatbeta_{AIPW2}(t)$}} \\
  \cmidrule{2-5}\cmidrule{7-10}
$t_1$ & 0.000 & 0.130 &  0.128  &  2.302&& 0.001 &  0.130  &  0.128 &     2.300\\
$t_2$ &  0.000 & 0.110 &  0.109 &   1.761&& 0.000 &  0.110 &  0.109 &     1.759\\
$t_3$ & -0.001 & 0.097 &  0.097 &   1.587&& -0.001 &  0.097 &  0.096 &     1.591\\
$t_4$ & 0.000 & 0.090 &  0.090&   1.389&& 0.000 &  0.090 &   0.090 &     1.389\\
$t_{end}$ & 0.000 &  0.090  & 0.090  &   1.123&& 0.000 &  0.090 &   0.090 &     1.123\\*[0.03in]
  \midrule
& \multicolumn{9}{@{}c@{}}{\textbf{(b) Alternative Hypothesis}} \\
  &\multicolumn{4}{@{}c@{}}{\textbf{$\hatbeta_{\Tf}(t)$}} &&\multicolumn{4}{@{}c@{}}{\textbf{ $\hatbeta_{IPW}(t)$}} \\
  \cmidrule{2-5}\cmidrule{6-9}
$t_1$ & -0.291 &  0.220  &  0.216 &    1.000&& -0.292 &  0.153  &  0.151 &     2.072\\
$t_2$ & -0.291 &  0.164&  0.162  &    1.000&& -0.291 &  0.130  &  0.129 &     1.591\\
$t_3$ & -0.292 &  0.136  &  0.135  &    1.000&& -0.291 &  0.115  &  0.114 &     1.412\\
$t_4$ & -0.291 &  0.119  &  0.118  &    1.000&& -0.290 &  0.107 &  0.107  &    1.242\\
$t_{end}$ & -0.290 &  0.107 &  0.107 &  1.000 && -0.290 &  0.107  &  0.107 &     1.000\\*[0.03in]
 &\multicolumn{4}{@{}c@{}}{\textbf{$\hatbeta_{AIPW1}(t)$}} &&\multicolumn{4}{@{}c@{}}{\textbf{$\hatbeta_{AIPW2}(t)$}} \\
  \cmidrule{2-5}\cmidrule{6-9}
$t_1$ & -0.291 &  0.147  &  0.146 &     2.246&& -0.291 &  0.147 &   0.145  &    2.249\\
$t_2$ & -0.291 &  0.124  &  0.123  &    1.734&& -0.291 &  0.124  &  0.123  &    1.736\\
$t_3$ &  -0.291 &  0.110  &  0.109 &     1.545&& -0.291 &  0.109  &  0.109 &     1.548\\
$t_4$ & -0.290 &  0.102  &  0.102 &    1.377&& -0.290 &  0.102  &  0.101&     1.373\\
$t_{end}$ & -0.290 &  0.101 &  0.101  &    1.109&& -0.290 &  0.101 &  0.101   &   1.109\\
        \bottomrule
\end{tabular*}
\end{table*}
\end{center}

\vspace*{0.15in}

\noindent{\em Simulation Scenario 3:  Continuous Outcome:}
Under the null hypothesis, based on 10000 Monte Carlo data sets, the
Monte Carlo sample covariance matrices of
$\{\hatbeta(t_1),\ldots,\hatbeta(t_4),\hatbeta(t_{end})\}$, where
$\hatbeta(t)$ is each of $\hatbeta_{\Tf}(t)$, $\hatbeta_{IPW}(t)$,
$\hatbeta_{AIPW1}(t)$, and $\hatbeta_{AIPW2}(t)$, are given by
\begin{verbatim}
betahat_Tf
      [,1] [,2] [,3] [,4] [,5]
[1,] 11.67 7.84 5.85 4.65 3.85
[2,]  7.84 7.89 5.87 4.69 3.90
[3,]  5.85 5.87 5.83 4.65 3.87
[4,]  4.65 4.69 4.65 4.63 3.86
[5,]  3.85 3.90 3.87 3.86 3.88

\betahat_IPW
      [,1] [,2] [,3] [,4] [,5]
[1,] 11.67 7.84 5.85 4.65 3.85
[2,]  7.84 7.89 5.87 4.69 3.90
[3,]  5.85 5.87 5.83 4.65 3.87
[4,]  4.65 4.69 4.65 4.63 3.86
[5,]  3.85 3.90 3.87 3.86 3.88

\betaht_AIPW1
      [,1] [,2] [,3] [,4] [,5]
[1,] 10.65 6.81 4.82 3.82 3.14
[2,]  6.81 7.08 5.04 3.86 3.19
[3,]  4.82 5.04 5.13 3.96 3.16
[4,]  3.82 3.86 3.96 3.94 3.16
[5,]  3.14 3.19 3.16 3.16 3.17

betahat_AIPW2
     [,1] [,2] [,3] [,4] [,5]
[1,] 7.92 5.47 4.27 3.50 3.17
[2,] 5.47 5.44 4.17 3.49 3.16
[3,] 4.27 4.17 4.15 3.48 3.16
[4,] 3.50 3.49 3.48 3.46 3.16
[5,] 3.17 3.16 3.16 3.16 3.17
\end{verbatim}
Under the alternative $\beta_A = 6.24$, the analogous Monte Carlo
sample covariance matrices are
\begin{verbatim}
betahat_Tf
      [,1] [,2] [,3] [,4] [,5]
[1,] 11.71 7.87 5.88 4.67 3.86
[2,]  7.87 7.92 5.90 4.71 3.91
[3,]  5.88 5.90 5.85 4.67 3.89
[4,]  4.67 4.71 4.67 4.65 3.88
[5,]  3.86 3.91 3.89 3.88 3.89
\end{verbatim}

\clearpage
\begin{verbatim}
betahat_IPW
      [,1] [,2] [,3] [,4] [,5]
[1,] 11.71 7.87 5.88 4.67 3.86
[2,]  7.87 7.92 5.90 4.71 3.91
[3,]  5.88 5.90 5.85 4.67 3.89
[4,]  4.67 4.71 4.67 4.65 3.88
[5,]  3.86 3.91 3.89 3.88 3.89

betahat_AIPW1
      [,1] [,2] [,3] [,4] [,5]
[1,] 10.69 6.84 4.84 3.83 3.15
[2,]  6.84 7.10 5.06 3.87 3.20
[3,]  4.84 5.06 5.15 3.97 3.18
[4,]  3.83 3.87 3.97 3.95 3.17
[5,]  3.15 3.20 3.18 3.17 3.18

betahat_AIPW2
     [,1] [,2] [,3] [,4] [,5]
[1,] 7.94 5.49 4.29 3.52 3.18
[2,] 5.49 5.46 4.18 3.51 3.17
[3,] 4.29 4.18 4.16 3.49 3.18
[4,] 3.52 3.51 3.49 3.47 3.18
[5,] 3.18 3.17 3.18 3.18 3.18
\end{verbatim}

Table~\ref{t:scenario3} presents performance of the estimators under
the null and alternative hypotheses.  
\begin{center}
\begin{table*}[ht]%
  \caption{For Scenario 3 with continuous outcome,
    performance of estimators for $\beta$ under (a) the null
    hypothesis $\beta=0$ and (b) the alternative
    $\beta = 6.24$ at each interim analysis time
    $(t_1,\ldots, t_4) = (104,130,156,182)$ days and at the final
    analysis at $t_{end} = 208$ days MC Mean is the mean of 10000 Monte
    Carlo estimates; MC SD is the Monte Carlo standard deviation, Ave
    MC SE is the mean of Monte Carlo standard errors, and MSE ratio is
    the ratio of Monte Carlo mean square error for the AIPW2 estimator
    divided by that for the indicated estimator.\label{t:scenario3}}
  \centering
  \begin{tabular*}{500pt}{@{\extracolsep\fill}lrcccp{0.1in}rccc@{\extracolsep\fill}}
  \toprule
    & \textbf{MC Mean} & \textbf{MC SD}  & \textbf{Ave MC SE}  & \textbf{MSE ratio}
& &\textbf{MC Mean} & \textbf{MC SD}  & \textbf{Ave MC SE}  & \textbf{MSE ratio}\\
& \multicolumn{9}{@{}c@{}}{\textbf{(a) Null Hypothesis}} \\
  &\multicolumn{4}{@{}c@{}}{\textbf{$\hatbeta_{\Tf}(t)$}} &&\multicolumn{4}{@{}c@{}}{\textbf{ $\hatbeta_{IPW}(t)$}} \\
  \cmidrule{2-5}\cmidrule{7-10}
$t_1$ & 0.012 & 3.416  & 3.415 &  1.000 && 0.012 & 3.416  & 3.380  & 1.000\\
$t_2$ & 0.010 & 2.809  & 2.781  & 1.000 && 0.010 & 2.809  & 2.763 &   1.000\\
$t_3$ &  -0.001 & 2.414  & 2.407 &  1.000 && -0.001 & 2.414  & 2.395 &   1.000\\
$t_4$ &  -0.005 & 2.152  & 2.151  &  1.000 && -0.005 & 2.152  & 2.142  &  1.000 \\
$t_{end}$  & 0.005 & 1.969  & 1.962  & 1.000 && 0.005 & 1.969  & 1.956  & 1.000 \\
  &\multicolumn{4}{@{}c@{}}{\textbf{$\hatbeta_{AIPW1}(t)$}} &&\multicolumn{4}{@{}c@{}}{\textbf{$\hatbeta_{AIPW2}(t)$}} \\
  \cmidrule{2-5}\cmidrule{7-10}
$t_1$ & 0.007 & 3.264  & 3.222 &     1.095 && -0.009 & 2.813  & 2.721   &   1.474\\
$t_2$ &  -0.006 & 2.660  & 2.609  &    1.115 &&  -0.003 & 2.332  & 2.284 &     1.452\\
$t_3$ & 0.002 & 2.265  & 2.247   &   1.136 && 0.001 & 2.037  & 2.014 &     1.405\\
$t_4$ & -0.001 & 1.985  & 1.975  &    1.176 &&  -0.001 & 1.859  & 1.839  &    1.340\\
$t_{end}$ & 0.008 & 1.780  & 1.772  &    1.225 && 0.008 & 1.780  & 1.772  &    1.225\\*[0.03in]
  \midrule
& \multicolumn{9}{@{}c@{}}{\textbf{(b) Alternative Hypothesis}} \\
  &\multicolumn{4}{@{}c@{}}{\textbf{$\hatbeta_{\Tf}(t)$}} &&\multicolumn{4}{@{}c@{}}{\textbf{ $\hatbeta_{IPW}(t)$}} \\
  \cmidrule{2-5}\cmidrule{6-9}
$t_1$ & 6.230 & 3.422  & 3.421  & 1.000 && 6.230 & 3.422  & 3.386  & 1.000 \\
$t_2$ & 6.208 & 2.815  & 2.786  & 1.000 && 6.208 & 2.815  & 2.768  & 1.000\\
$t_3$ & 6.216 & 2.419  & 2.411  & 1.000 && 6.216 & 2.419  & 2.399 & 1.000\\
$t_4$ & 6.213 & 2.157  & 2.154  & 1.000 && 6.213 & 2.157  & 2.146  & 1.000\\
$t_{end}$ & 6.223 & 1.973  & 1.966 & 1.000 && 6.223 & 1.973  & 1.959 & 1.000\\ 
 &\multicolumn{4}{@{}c@{}}{\textbf{$\hatbeta_{AIPW1}(t)$}} &&\multicolumn{4}{@{}c@{}}{\textbf{$\hatbeta_{AIPW2}(t)$}} \\
  \cmidrule{2-5}\cmidrule{6-9}
$t_1$ & 6.225 & 3.269  & 3.227  &    1.096 && 6.210 & 2.818  & 2.726  &    1.474\\
$t_2$ & 6.212 & 2.665  & 2.613  &    1.115 && 6.216 & 2.336  & 2.288  &    1.452\\
$t_3$ & 6.220 & 2.269  & 2.250  &    1.137 && 6.219 & 2.041  & 2.017  &    1.405\\
$t_4$ & 6.217 & 1.989 &  1.979  &    1.176 && 6.217 & 1.863  & 1.842  &    1.341\\
$t_{end}$ & 6.226 & 1.783 &  1.775  &    1.225 && 6.226 & 1.783  & 1.775   &   1.225\\
        \bottomrule
\end{tabular*}
\end{table*}
\end{center}

\clearpage
\bibliography{timelag}%

\end{document}